\documentclass[a4paper,UKenglish,cleveref, autoref, thm-restate]{lipics-v2021}

\hideLIPIcs
\nolinenumbers

\title{Beyond Optimality: Hyperoptimal Strategies in Concurrent Games} 

\author{Ali Asadi}{Institute of Science and Technology Austria}{ali.asadi@ista.ac.at}{https://orcid.org/0009-0005-2839-953X}{}

\author{Léonard Brice}{Institute of Science and Technology Austria}{leonard.brice@ista.ac.at}{https://orcid.org/0000-0001-7748-7716}{}

\author{Thomas A. Henzinger}{Institute of Science and Technology Austria}{tah@ista.ac.at}{https://orcid.org/0000-0002-2985-7724}{This work is a part of project VAMOS that has received funding from the European Research Council (ERC), grant agreement No 101020093.
 Part of this work was realised when the first author was an FNRS aspirant at Université libre de Bruxelles.}

\authorrunning{A. Asadi, L. Brice, and T. A. Henzinger} 

\Copyright{A. Asadi, L. Brice, and T. A. Henzinger} 



\category{} 

\relatedversion{} 

\EventEditors{Manuel Bodirsky and Ichiro Hasuo}
\EventNoEds{55}
\EventLongTitle{55th International Colloquium on Automata, Languages, and Programming}
\EventShortTitle{ICALP 2027}
\EventAcronym{ICALP}
\EventYear{2027}
\EventDate{July 5--9, 2027}
\EventLocation{Krak\'{o}w, Poland}
\EventLogo{}
\SeriesVolume{55}
\ArticleNo{}

\usepackage{graphicx}
\usepackage{array}
\usepackage{tikz}
\usepackage[new]{old-arrows}
\usepackage{algpseudocode}
\usepackage{algorithm}

\usetikzlibrary{automata, arrows, positioning, decorations.markings, decorations.pathreplacing}
\usetikzlibrary {decorations.pathmorphing, decorations.shapes}
\usetikzlibrary{backgrounds,fit,shapes.geometric}
\usetikzlibrary {graphs, quotes}
\usetikzlibrary{calc}
\usetikzlibrary{shapes,shapes.geometric,fit}
\usetikzlibrary {graphs,shapes.geometric,quotes}
\usetikzlibrary{decorations.pathreplacing}
\usetikzlibrary{automata}
\usetikzlibrary{calc}
\usetikzlibrary{arrows,automata,decorations.pathmorphing}
\usetikzlibrary{shapes,shapes.geometric,fit,calc,automata}

\newcommand{\Dist}{\mathsf{Dist}}
\newcommand{\Plays}{\mathsf{Plays}}
\newcommand{\Hist}{\mathsf{Hist}}
\newcommand{\Supp}{\mathsf{Supp}}
\newcommand{\first}{\mathsf{first}}
\newcommand{\last}{\mathsf{last}}
\newcommand{\Strat}{\mathsf{Strat}}
\newcommand{\Abs}{\mathsf{Abs}}
\renewcommand{\Loop}{\mathsf{Loop}}
\newcommand{\abs}{\mathsf{abs}}
\newcommand{\imm}{\mathsf{imm}}
\newcommand{\cons}{\mathsf{cons}}
\newcommand{\crit}{\mathsf{crit}}
\newcommand{\stall}{\mathsf{stall}}

\renewcommand{\d}{\mathsf{d}}

\newcommand{\hide}{\mathsf{H}}
\newcommand{\run}{\mathsf{R}}
\newcommand{\wait}{\mathsf{W}}
\newcommand{\throw}{\mathsf{T}}

\newcommand{\PSPACE}{\mathsf{PSPACE}}

\renewcommand{\phi}{\varphi}
\renewcommand{\epsilon}{\varepsilon}

\renewcommand{\Game}{\mathcal{G}}

\newcommand{\Ada}{\mathfrak{A}}
\newcommand{\Bertrand}{\mathfrak{B}}

\newcommand{\prob}{\mathbb{P}}
\newcommand{\expec}{\mathbb{E}}

\newcommand{\RR}{\mathbb{R}}

\newcommand{\QQ}{\mathbb{Q}}

\tikzset{every node/.append style={initial text={}}}
\tikzset{circlev/.style={state,minimum size=20pt,fill, top color=white, bottom color=black!20}}

\Crefname{clm}{Claim}{Claims}

 \usepackage{todonotes}

\usepackage[framemethod=TikZ]{mdframed}

\newcommand{\subp}[1]{\subparagraph*{#1.}}

\begin{document}

\maketitle

\begin{abstract}
    We study the topology of optimality in concurrent games.
    
    We consider two-player zero-sum concurrent games played on graphs, where the players' payoffs depend on the absorbing state that is reached---also called recursive games.
    In such games, it is known that the max-min value equals the min-max value: we simply call it value.
    Then, optimal strategies are defined as strategies that guarantee at least the value, whatever the opponent does.

    We introduce hyperoptimal strategies, i.e., strategies that guarantee strictly more than the value---although the opponent may get arbitrarily close to it.
    The existence of a hyperoptimal strategy implies that the opponent has no optimal strategy, but we show that the converse is false.

    We provide a characterisation of games where a hyperoptimal strategy exists, in two game classes: absorbing games, i.e. games with only one non-absorbing state, and negative recursive games, i.e. games where all rewards are non-positive---with safety games as a special case.
    We show that in both cases, when a hyperoptimal strategy exists, there is one obtained by a statewise private mixture of two optimal memoryless strategies: an independent private fair coin selects between the mixed actions they prescribe at each non-absorbing state at the beginning of the game.
    As an application, we show that deciding whether a strategy guarantees an expected payoff strictly larger than a given rational threshold is solvable in polynomial time for absorbing games. For negative recursive games, this problem belongs to $\exists\RR\subseteq\PSPACE$ and is square-root-sum-hard, already for safety games.
\end{abstract}

\subp{AI disclosure}
AI tools were used to suggest proof ideas. The proofs exposed here were however fully written by the authors, and those remain fully responsible of their content.

\section{Introduction}\label{sec:intro}
Game theory provides a powerful toolbox to study \emph{open systems}, i.e., computer-controlled systems that interact with an unpredictable environment.
Such systems are modelled as a game, where a player, here called \emph{Ada}, intends to maximise a payoff function or to satisfy a Boolean objective; she faces a second player, here called \emph{Bertrand}, conservatively assumed to act in a fully adversarial way.
Historically, \emph{turn-based games}, where Ada and Bertrand play asynchronously, have received more attention, since they offer convenient properties.
However, \emph{concurrent games}, where they may have to select actions simultaneously, are more suited to model synchronous systems.

\subp{The model}
Throughout this paper, our model is that of \emph{concurrent games played on graphs with terminal rewards}, also known as \emph{simple quantitative games} or \emph{recursive games}.
Such a game is made of \emph{states}, from which both Ada and Bertrand have to select an \emph{action}: then, each action pair entails a probability distribution over the next state the game transitions to, from which the same process starts again.
Some states are \emph{absorbing}, meaning that once they are reached, no action pair can make the game transition to another state.
Ada's payoff depends on the absorbing state that is reached; if none is, her payoff is $0$.
As specific cases, we consider:
\begin{itemize}
    \item \emph{absorbing} games, where only one state is non-absorbing;

    \item \emph{positive} (resp. \emph{negative}) recursive games, where all rewards are non-negative (resp. non-positive)\footnote{Positive recursive games are more standard than the dual notion. However, in such games, only the antagonist may have hyperoptimal strategies. Since those are the focus of this paper, we mostly study negative recursive games.};

    \item \emph{reachability} (resp. \emph{safety}) games, where all rewards are $0$ and $1$ (resp. $0$ and $-1$).
\end{itemize}

\subp{Qualitative problems}
In such games, most works have been dedicated to \emph{qualitative problems}.
Ada is then given a Boolean objective, and we may ask three questions, each being a relaxation of the previous one: Given a game,
\begin{enumerate}
    \item \emph{Sure problem:} does Ada have a strategy that guarantees the objective against every strategy of Bertrand?
    
    \item \emph{Almost-sure problem:} does Ada have a strategy such that for every strategy of Bertrand, the objective is satisfied with probability 1?
    
    \item \emph{Limit-sure problem:} is it the case that for every $\epsilon > 0$, Ada has a strategy such that for every strategy of Bertrand, the objective is satisfied with probability at least $1-\epsilon$?
\end{enumerate}

\subp{Value}
When Ada is instead given a payoff function to maximise, the focus switches to the maximal payoff Ada can guarantee against all Bertrand's strategies, or more precisely the supremum over Ada's strategies of the infima over Bertrand's strategies of Ada's expected payoffs.
That quantity is called the \emph{max-min value} of the game\footnote{This term is widely stabilised in the literature, but \emph{sup-inf value} would actually be more accurate, since there is no general guarantee that the supremum and the infima are attained.}, or simply \emph{value} when the game is \emph{determined}, i.e. when it equals its dual value computed from Bertrand's perspective, which is the case in the models we consider~\cite{Everett1957Recursive}.


\subp{Three shades of optimality}
Nevertheless, computing the value of a game does not tell the whole story.
First, because Ada may or may not have an \emph{optimal strategy}, i.e., a strategy that actually secures the value, and realises the supremum in its definition---this is the quantitative equivalent of the difference between almost-sure winning and limit-sure winning.

Second, because even when Ada has an optimal strategy, Bertrand may or may not have a \emph{best response} to that strategy, i.e., a response that confines her expected payoff to the value, realising the infimum in its definition.
An optimal strategy with no best response then secures strictly more than the value, since against such a strategy, Bertrand can get arbitrarily close to the value but cannot reach it.
Such strategies have not been studied systematically in the literature.
We call them \emph{hyperoptimal strategies}.

With the classical notion of \emph{$\epsilon$-optimal strategies} (strategies for Ada such that for every Betrand's strategy, Ada's expected payoff is at least $v - \epsilon$, where $v$ is the value), optimal strategies and hyperoptimal strategies form three levels of requirement one can have about a strategies, whose existence conditions differ:
\begin{itemize}
    \item for every $\epsilon > 0$, in every recursive game, we always have existence of $\epsilon$-strategies;

    \item optimal strategies are guaranteed to exist in negative recursive games~\cite{DEALFARO2004374}, and may or may not exist in positive recursive games;

    \item hyperoptimal strategies may or may not exist in negative recursive games (see the example below), but are guaranteed \emph{not} to exist in positive recursive games (where Bertrand has an optimal strategy by the above).
\end{itemize}

\subp{Quantitative problems}
When it comes to define decision problems in a quantitative setting, there are at least four different questions that one may ask:
Given a game and a threshold $\alpha$,

\begin{enumerate}
    \item\label{it:strict_val} \emph{Strict value problem:} Is the value strictly larger than $\alpha$?

    \item\label{it:strict_strat} \emph{Strict strategy problem:} Does Ada have a strategy such that for every strategy of Bertrand, her expected payoff is strictly larger than $\alpha$?
    
    \item\label{it:ns_strat} \emph{Non-strict strategy problem:} Does Ada have a strategy such that for every strategy of Bertrand, her expected payoff is at least $\alpha$?
    
    \item\label{it:ns_val} \emph{Non-strict value problem:} Is the value at least $\alpha$?
\end{enumerate}

Note that, when $\alpha = 1$, the strict problems are trivial, the non-strict strategy problem coincides with the almost-sure problem, and the non-strict value problem with the limit-sure problem.
When $\alpha$ is the value, the non-strict strategy problem amounts to asking whether Ada has an optimal strategy (the \emph{optimality problem}), and the strict strategy problem whether she has a hyperoptimal one (the \emph{hyperoptimality problem}).

\begin{figure}
    \centering
    \begin{tikzpicture}[node distance = 3.5cm]
        \node[state, rectangle, rounded corners, fill, top color=white, bottom color=gray!30] (1) {Strict value problem};
        \node[state, rectangle, rounded corners, below of=1, fill, top color=white, bottom color=gray!30] (2) {Strict strategy problem};
        \node[state, rectangle, rounded corners, right of=2, fill, top color=white, bottom color=gray!30, align=left] (3) {Non-strict\\strategy problem};
        \node[state, rectangle, rounded corners, above of=3, fill, top color=white, bottom color=gray!30, align=left] (4) {Non-strict\\value problem};

        \path[<->] (1) edge[bend right] node[left, align=right] {equivalent if hyperoptimal\\ strategies do not exist\\(e.g. in positive\\recursive games)} (2);
        \path[<->] (3) edge[bend right] node[right, align=left] {equivalent if optimal\\ strategies exist\\(e.g. in negative\\recursive games)} (4);
        \path[<->] (1) edge[bend right] node[below] {duals} (4);
        \path[<->, dashed] (2) edge[bend left] node[above] {not duals} (3);
    \end{tikzpicture}
    \caption{Links between the four quantitative problems}
    \label{fig:links_problems}
\end{figure}
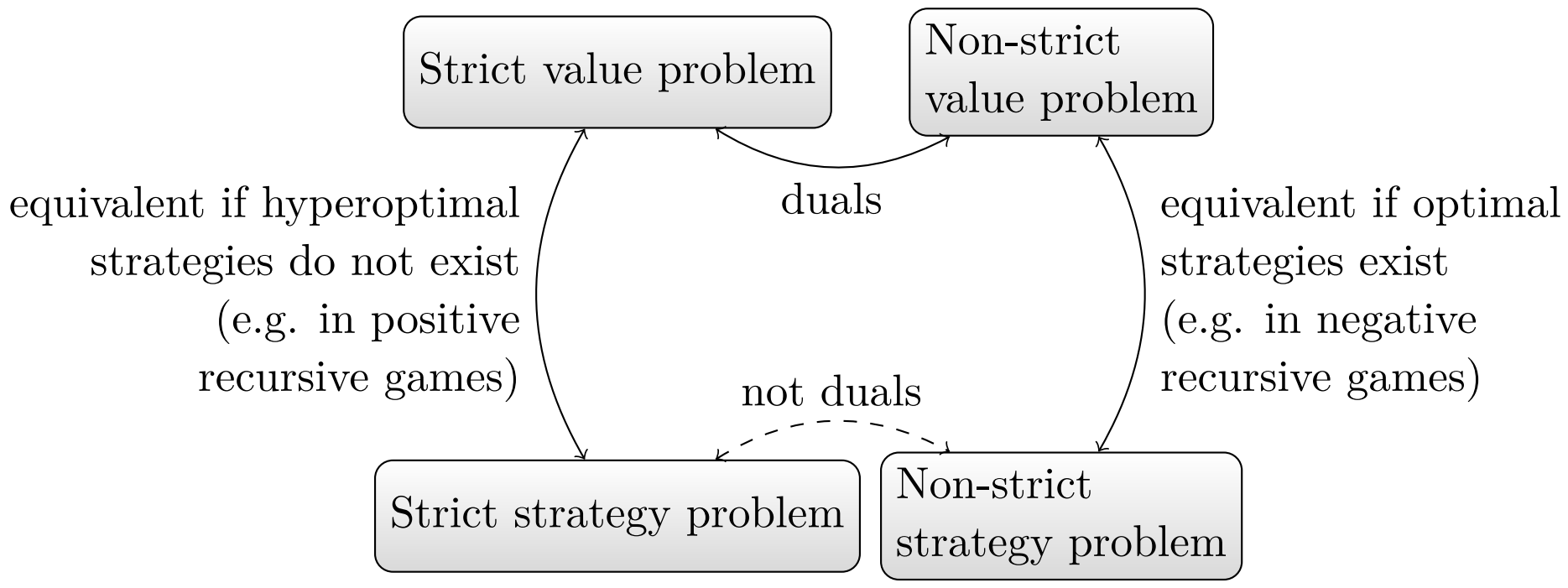

In the literature, the distinction between these four problems has often been overlooked.
A reason may be that, in many cases, solving one enables to solve others, either by equivalence or by duality: these links are exposed in \Cref{fig:links_problems}.
As for duality, the value problems are duals: a game has value larger than $\alpha$ if and only if the same game, where Ada and Bertrand are swapped and payoffs negated, does not have value larger than or equal to $-\alpha$.

As for equivalence, first note that each problem is a relaxation of the previous one: we have \ref{it:strict_val} $\Rightarrow$ \ref{it:strict_strat} $\Rightarrow$ \ref{it:ns_strat} $\Rightarrow$ \ref{it:ns_val}.
The implication \ref{it:strict_strat} $\Rightarrow$ \ref{it:ns_strat} is strict in every non-trivial class of games, but implication \ref{it:ns_strat} $\Rightarrow$ \ref{it:ns_val} turns out to be an equivalence in every class where Ada is guaranteed to have an optimal strategy, that is, a strategy that guarantees at least the value of the game against every strategy of Bertrand.
Symmetrically, the implication \ref{it:strict_val} $\Rightarrow$ \ref{it:strict_strat} is an equivalence as soon as it is guaranteed that Ada has no hyperoptimal strategy.

\begin{figure}[h] 
			\centering
            \begin{subfigure}[b]{0.45\textwidth}
        			 \centering
                    \begin{tabular}{c|c|c|}
                        & $\hide$ & $\run$ \\
                        \hline
                        $\wait$ & $\Loop$ & $-1$ \\
                        \hline
                        $\throw$ & $-1$ & $0$ \\
                        \hline
                    \end{tabular}
                    \caption{The table view}
                    \label{fig:hide-or-run-table}
            \end{subfigure}
            \hfill
            \begin{subfigure}[b]{0.45\textwidth}
			    \centering
    \begin{tikzpicture}[node distance = 2cm, ->]
        \node[circlev, initial above] (s0) {$s_0$};
        \node[circlev, above right of=s0] (ssafe) {$-1$};
        \node[circlev, below right of=s0] (swet) {$0$};

        \path (s0) edge node[below right] {$\wait\run, \throw\hide$} (ssafe);
        \path (s0) edge node[below left] {$\throw\run$} (swet);
        \path (s0) edge[loop left] node[left] {$\wait\hide$} (s0);
        \path (ssafe) edge[loop right] node[right] {$*$} (ssafe);
        \path (swet) edge[loop right] node[right] {$*$} (swet);
    \end{tikzpicture}
                    \caption{The graph view}
                    \label{fig:hide-or-run-graph}
            \end{subfigure}
        \caption{The hide-or-run game}
    \label{fig:hide-or-run}
\end{figure}
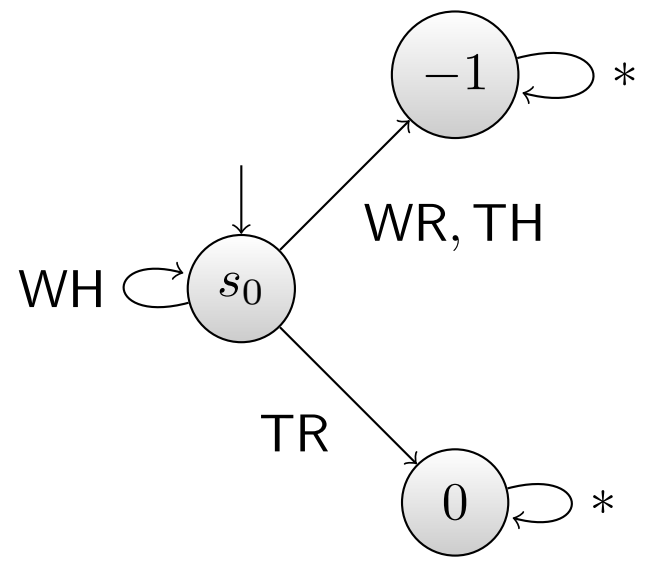

\subp{An example}
Consider the safety game depicted by \Cref{fig:hide-or-run}.
Ada is given a snowball, and can at each round choose to throw it or to wait.
Bertrand, on his side, hides behind a rock: at each round, he can keep hiding, or run to his house.
Ada's goal is to prevent him from reaching the house dry.
Note that this game is deterministic: each action pair fully determines the next state.

This game is often presented from Bertrand's perspective, as a reachability game.
It is well-known that Bertrand can limit-surely reach home safe, by running with probability $\epsilon$ at each round: the value of this game is therefore $-1$.
However, he cannot reach home safe almost surely---in other words, he has no optimal strategy.

But there is more: there exists a strategy for Ada such that for every strategy of Bertrand, the probability that Bertrand gets home safe is less than 1.
Indeed, consider the strategy $\sigma$ for Ada that, at round $n = 0, 1, \dots$, throws with probability $2^{-n-2}$.
Against $\sigma$, any strategy for Bertrand that eventually runs with positive probability implies a positive probability of being hit by the ball.
And if Bertrand never runs, then with probability $\prod_n (1-2^{-n-2}) \approx 0.578$, Ada never throws, and the game stays in state $s_0$ forever.
Thus, even though this game's value is $-1$, Ada has a strategy against which Bertrand can approach this value but never reach it: a hyperoptimal strategy.

Let us note that this hyperoptimal strategy uses infinite memory: this is necessary, since responding to a finite-memory strategy amounts to play in a finite Markov decision process, where optimal strategies exist---therefore, there exists a best response to every finite-memory strategy.
On the other hand, we can already note that this strategy can be interpreted in the following way: Ada secretly tosses a coin, and depending on the outcome, either never throws, or throws at each round with probability $0.5$.







\subp{Optimality and hyperoptimality are not duals}
Trivially, whenever Bertrand has an optimal strategy, Ada has no hyperoptimal strategy.
The hide-and-throw example suggests that the converse is true.
This may explain why hyperoptimality has not been defined as such up to date: it would simply be another way to say that the opponent has no optimal strategy, and Problems~\ref{it:strict_strat} and~\ref{it:ns_strat} would be duals.
However, this natural conjecture is false.

\begin{figure}[h] 
			\centering
            \begin{subfigure}[b]{0.45\textwidth}
        			 \centering
                    \begin{tabular}{c|c|c|}
                        & $\hide$ & $\run$ \\
                        \hline
                        $\wait$ & $\Loop$ & $-1$ \\
                        \hline
                        $\throw$ & $-0.5$ & $0$ \\
                        \hline
                    \end{tabular}
                    \caption{The table view}
                    \label{fig:explosive-table}
            \end{subfigure}
            \hfill
            \begin{subfigure}[b]{0.45\textwidth}
			    \centering
    \begin{tikzpicture}[node distance = 2cm, ->]
        \node[circlev, initial above] (s0) {$s_0$};
        \node[circlev, above right of=s0] (ssafe) {$-1$};
        \node[circlev, below right of=s0] (swet) {$0$};
        \node[circlev, above right of=swet] (shalf) {$-0.5$};

        \path (s0) edge node[above left] {$\wait\run$} (ssafe);
        \path (s0) edge node[below left] {$\throw\run$} (swet);
        \path (s0) edge node[above] {$\throw\hide$} (shalf);
        \path (s0) edge[loop left] node[left] {$\wait\hide$} (s0);
        \path (ssafe) edge[loop right] node[right] {$*$} (ssafe);
        \path (swet) edge[loop right] node[right] {$*$} (swet);
        \path (shalf) edge[loop right] node[right] {$*$} (shalf);
    \end{tikzpicture}
                    \caption{The graph view}
                    \label{fig:explosive-graph}
            \end{subfigure}
        \caption{The hide-or-run game with explosive snowballs}
    \label{fig:explosive}
\end{figure}
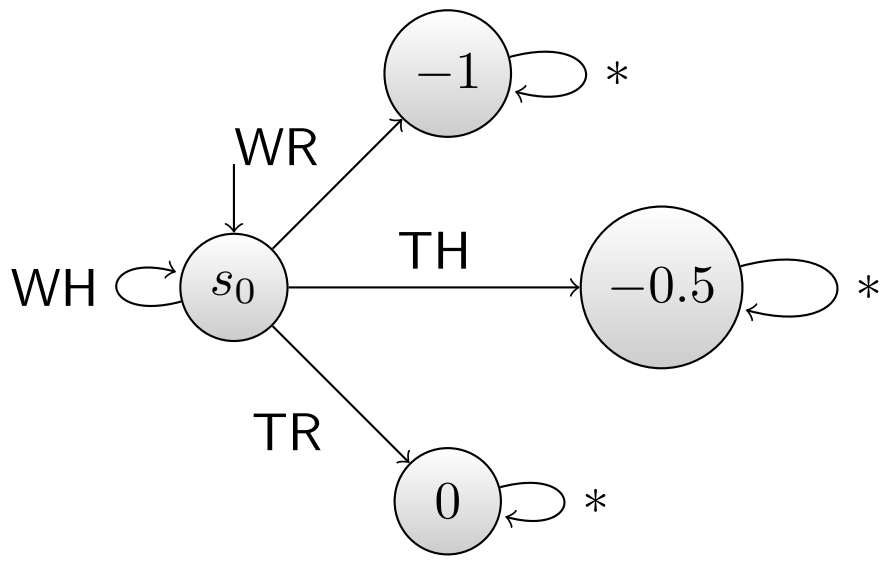

Consider the hide-or-run game, but where the snowballs are now explosive: if Ada throws when Bertrand is hiding, Bertrand is partially hit, and Ada's payoff is $-0.5$ instead of $-1$.
The value of this game is now $-0.5$, and Bertrand has, again, no optimal strategy.\footnote{
Indeed, if he never runs, then Ada can respond by never throwing, and the payoff is then $0 > -0.5$.
Otherwise, Ada can respond by waiting, and then throwing at the first round at which he runs with probability $p > 0$: then, the expected payoff is $(1-p) \times (-0.5) > -0.5$.}
However, Ada has no hyperoptimal strategy in this game.\footnote{Consider an optimal strategy $\sigma$ for Ada, and let $p_n$ be the probability that Ada throws at round $n = 1, 2, \dots$---conditioned on still being in $s_0$ at that round.
If we have $p_n > 0.5$ for all $n$, then Bertrand can respond by always hiding: then, it is almost sure that Ada eventually throws, giving payoff $-0.5$.
If there is some $n$ with $p_n \leq 0.5$, then Bertrand can respond by hiding up to round $n$ and then running, giving expected payoff $(1-p_n) \times (-1) \leq -0.5$.
The strategy $\sigma$ is therefore not guaranteeing an expected payoff strictly above $-0.5$, and is thus not hyperoptimal.}

Consequently, the hyperoptimality problem is not the dual of the optimality problem; and more generally, the strict strategy problem is not the dual of the non-strict one.
Outside of the realm of positive recursive games, the strict strategy problem is a genuinely open decision problem, which justifies the study of hyperoptimal strategies.

\subsection{Contribution and technical overview}

Our first contribution is the concept of hyperoptimality itself, and its relationship with other standard notions.
In particular, we exhibit a game (\Cref{fig:explosive}) where Bertrand does not have an optimal strategy and Ada does not have a hyperoptimal strategy, showing that the optimality and the hyperoptimality problems are not duals.

We then give a characterisation of games where Ada has a hyperoptimal strategy, in two orthogonal classes of recursive games: absorbing recursive games (with arbitrary rewards) and negative recursive games.
Our key result is the following: in every game in these two classes, if a hyperoptimal strategy exists, there is one that is obtained as a \emph{(statewise) mixture} of two memoryless optimal strategies---i.e., Ada secretly tosses a coin to decide which of the two strategies she follows in each state.
We prove that result separately in absorbing games and in negative recursive games.
The general case of recursive games is left open.

\subp{Absorbing games (\Cref{sec:absorbing})}
In absorbing games, proofs are simpler, and should help the reader understand the proofs of \Cref{sec:negative}.
We start by defining a few notations: for every Ada mixed action $x$ and Bertrand action $b$, we denote by $\abs(x, b)$ the probability of reaching an absorbing state at the next round if Ada follows $x$ and Bertrand plays $b$ (the \emph{absorption rate}), and by $\imm(x, b)$ the expected value of the reward obtained at the next round, viewing a loop on $s_0$ as a reward $0$.
We then define the polytope $P$ of \emph{positive} mixed actions for Ada as the set of mixed actions $x$ satisfying $\cons_b(x) \geq 0$ for every $b$, where $\cons_b: x \mapsto \imm(x, b) - v \cdot \abs(x, b)$.
We prove that the polytope $P$ is nonempty (\Cref{lm:positive_nonempty}), and that every optimal memoryless strategy is such that the corresponding mixed action is positive.
When the value $v$ is negative, we prove that the converse is true (\Cref{lm:positive_strategies}).

When the value $v$ is non-negative, we prove that Ada never has a hyperoptimal strategy (\Cref{thm:abs_v_nonneg}).
We therefore assume $v < 0$ in the following.
We consider a mixed action $p^\circ$ in the relative interior of $P$---and which therefore has maximal support among positive mixed actions.
We define \emph{critical actions} as actions $b$ for Bertrand such that $\cons_b(p^\circ) = 0$, or equivalently such that $\cons_b$ is zero on all $P$ (\Cref{lm:critical_relint}).
We show that the set of critical actions is always nonempty (\Cref{lm:critical_nonempty}), and that any best response to $p^\circ$ necessarily plays only critical actions (\Cref{lm:relint_best_response}).

We then introduce \emph{stalling actions} for Ada, as actions that guarantee looping on $s_0$ against all critical actions of Bertrand, and \emph{stalling strategies} as memoryless optimal strategies playing only stalling actions.
Then we show that Ada has a hyperoptimal strategy if and only if she has a stalling one.
Indeed, if she has a stalling strategy $p$, then she can play the strategy $\sigma$ obtained by tossing a secret coin to choose between $p$ and $p^\circ$.
Note that the support of $p$ is included in the support of $p^\circ$, hence if Ada follows $p$, Bertrand can never know it.
Any best response of Bertrand to $\sigma$ then has to be a best response to both $p$ and $p^\circ$; but a best response to $p^\circ$ can only play critical actions by \Cref{lm:relint_best_response}, and any strategy playing only critical actions against $p$ results in looping on $s_0$, and then  in a payoff $0$, which is strictly more than $v$.

\subp{Negative recursive games (\Cref{sec:negative})}
The result on negative recursive games follows the same intuition but requires additional steps.
First, \emph{positive} mixed actions can now be defined, in any state $s$, as mixed actions that constitute optimal strategies in the matrix game where Ada intends to maximise the value of the state reached at the next round.
Again, the set of positive mixed actions from each state $s$ is a nonempty polytope $P_s$ (\Cref{lm:negative_positive_nonempty}).
Strategies that always play positive mixed actions are called \emph{locally optimal}, and we use the standard result that in negative recursive games, locally optimal strategies are always optimal (\Cref{lm:negative_positive_optimal}).

Again, we define a memoryless strategy $p^\circ$ such that for every $s$, the mixed action $s$ lies in the relative interior of $P_s$, and therefore has maximal support among positive actions.
Critical actions for Bob at $s$ are best responses to $p^\circ(s)$ in the matrix game described above, and again, every best response to $p^\circ$ in the whole game must play only critical actions (\Cref{lm:negative_relint_best_response}).

As an equivalent to stalling strategies, we design a memoryless strategy for Ada, when she has a hyperoptimal strategy, that traps the game into a restricted region against all strategies of Bertrand that are best responses to $p^\circ$.
To do so, we show that Ada has \emph{no} hyperoptimal strategy if and only if Bertrand has a strategy playing critical actions that forces, against all locally optimal strategies of Ada, that almost surely either an absorbing state or a state of value $0$ is reached---in other words, if Bertrand has an almost surely winning strategy in a modified version of the game where he is constrained to play critical actions and Ada is constrained to play positive mixed actions, and where his objective is to reach those states (\Cref{thm:negative_statewise_mixture}).
When Bertrand does not have such a strategy, we use classical properties of reachability games to build a memoryless strategy for Ada that we can mix with $p^\circ$ to, again, obtain a hyperoptimal strategy.

\subp{Complexity of the strict strategy problem (\Cref{sec:complexity})}
We conclude with the complexity of the strict strategy problem (Problem~\ref{it:strict_strat}).
In absorbing games, by the above, we have a positive instance of the strict strategy problem if and only if we have either a memoryless strategy that guarantees strictly more than the threshold, or a memoryless strategy that guarantees the threshold and a stalling strategy.
This can be checked in polynomial time by linear programming.
In negative recursive games, there is a strategy guaranteeing strictly more than a specified threshold if and only if there is one that can be obtained as a mixture of two memoryless strategies.
The problem can then be rewritten as a sentence in the existential theory of the reals, which is decidable in $\PSPACE$.
We also show that the problem is at least as hard as the square-root-sum problem.

\subp{A quick discussion on these results}
The experienced reader may note that the complexity results obtained match exactly those known for the non-strict strategy problem.
This might come as a surprise: as we have seen, Ada can use memoryless strategies to ensure non-strict thresholds, but may require infinite memory to ensure strict ones.
However, what we have shown is that this necessity of infinite memory is an artifact of the behavioral view on strategies: by the mixture view, we obtain a simple shape for hyperoptimal strategies.

A question that we leave open, however, is whether such a result can be extended to general recursive games.
There, one new layer of difficulty comes from the fact that locally optimal strategies are no longer guaranteed to be optimal.
Moreover, plays where no absorbing state is reached and where Ada gets the payoff $0$ are perceived differently, depending on the value of the current state: Ada may seek it when her value is negative, and then, after an unfortunate event, suddenly try to avoid it.
Significantly new techniques may then need to be mobilised to adress this case.

\subsection{Related works}

The setting of concurrent games played on graphs, with reachability objectives, was introduced by de Alfaro, Henzinger, and Kupferman~\cite{DEALFARO2007188}.
That paper defines the qualitative theory we rely on, starting from the difference between sure winning, almost-sure winning, and limit-sure winning, and the standard hide-or-run example where the safety player has a hyperoptimal strategy.
The qualitative theory was subsequently extended to $\omega$-regular objectives~\cite{alfaro2002, 10.1145/1970398.1970404}.
Algorithms and complexity bounds for the quantitative analysis of concurrent stochastic games with parity objectives are studied in~\cite{DBLP:conf/fsttcs/AsadiCSS24}.

\emph{Recursive games}, under this name, are studied in a slightly different community.
The existence of a value in every recursive game is Everett's theorem~\cite{Everett1957Recursive}, which is the undiscounted counterpart of Shapley's result on discounted-sum games~\cite{Shapley1953StochasticGames}.
For a survey of the classical theory, see~\cite{Sorin02}.
Many researchers have focused on computing the value of a (positive) recursive game, with work on exact algorithms~\cite{DBLP:conf/stoc/HansenKLMT11,OliuBarton2021}, strategy improvement~\cite{DBLP:journals/jcss/ChatterjeeAH13}, termination criteria~\cite{DBLP:conf/soda/ChatterjeeAH09}, or value iteration~\cite{DBLP:journals/mst/HansenIM14}.
In positive recursive games, there always are $\epsilon$-optimal memoryless strategies~\cite{bordais_et_al:LIPIcs.CSL.2022.7}, and comparing the value to a given rational threshold (Problems~\ref{it:strict_val} or~\ref{it:ns_val} above) lies in the class $\exists\RR \subseteq \PSPACE$, and is hard for the square-root-sum problem~\cite{10.1007/11787006_28}.
That hardness is no accident: the values of such games need not be rational~\cite{OliuBartonVigeral2023}, and in fact every real algebraic number is the value of some absorbing recursive game~\cite{AsadiChatterjee2026}.

Finally, let us note that our threshold problem is one of the problems that must be resolved to prove the decidability of model-checking for the Probabilistic ATL logic on concurrent games~\cite{10.1007/978-3-642-28756-5_22}, and of its extension Individually Randomised ATL~\cite{10.1007/978-3-032-32519-8_12}, a currently open question to which this paper contributes.

\section{Definitions}\label{sec:background}
\subsection{Probabilities}

For a finite set $X$, we denote by $\Dist(X)$ the set of probability distributions over $X$, i.e. the set of mappings $d: X \rightarrow [0,1]$ satisfying $\sum_{x \in X} d(x) = 1$.
We denote the support of $d$ as $\Supp(d) = \{x \in X \mid d(x) > 0\}$.
Given $x \in X$, the associated \emph{Dirac} distribution is the distribution that maps $x$ to $1$, and every $y \in X \setminus \{x\}$ to $0$; we denote it by $\d_x$.

\subsection{Games}

\begin{definition}[Two-player concurrent game] \label{def:game}
    A \emph{two-player concurrent game} is a tuple $\Game = (S, s_0, A, B, \delta, \mu)$, consisting of:
    \begin{itemize}
        \item a set $S$ of \emph{states};
        
        \item an \emph{initial} state $s_0 \in S$;

        \item a set $A$ of \emph{actions} for Ada;

        \item a set $B$ of actions for Bertrand;

        \item a \emph{transition function} $\delta: S \times A \times B \to \Dist(S)$;

        \item a \emph{payoff function} $\mu: S^\omega \to \RR$.
    \end{itemize}
\end{definition}

\begin{definition}[Play, history]
    A \emph{play} in $\Game$ is an infinite sequence $\pi = (s_0, a_0, b_0, s_1, \dots) \in (S \cdot A \cdot B)^\omega$.
    A \emph{history} in $\Game$ is a finite sequence $h = (s_0, a_0, b_0, s_1, \dots, s_n) \in (S \cdot A \cdot B)^* S$.
\end{definition}

We call $n$ the \emph{length} of $h$, and denote it by $|h|$.
Given a play $\pi$ (resp. a history $h$), we denote the first state of $\pi$ (resp. $h$) by $\first(\pi)$ (resp. $\first(h)$).
Given a history $h$, we denote the last state of $h$ by $\last(h)$.
Given a play $\pi = (s_0, a_0, b_0, s_1, \dots)$, we abusively write $\mu(\pi)$ for the payoff $\mu(s_0 s_1 \dots)$.
We denote the set of plays (resp. histories) in the game $\Game$ by $\Plays(\Game)$ (resp. $\Hist(\Game)$).

A vertex $s$ is \emph{controlled} by Ada if for every Ada action $a$ and every two Bertrand actions $b, b'$, we have $\delta(s, a, b) = \delta(s, a, b')$; the definition of vertices controlled by Bertrand is symmetric.
A game is \emph{turn-based} if every vertex is controlled by either Ada or Bertrand.

A vertex $s$ is \emph{absorbing} if for every pair $(a, b)$, we have $\delta(s, a, b) = s$.
A \emph{recursive game} is a game where for each play $\pi$, the payoff $\mu(\pi)$ is $0$ if the play $\pi$ contains no absorbing state, and otherwise depends only on the first absorbing state the play $\pi$ stabilises in.
Given an absorbing state $s$, we write $\mu(s)$ for the payoff of every play that stabilises in $s$.
An \emph{absorbing game} is a recursive game where only the initial state $s_0$ is not absorbing.
A recursive game is \emph{negative} (resp. \emph{positive}) if all payoffs are non-positive (resp. non-negative).
It is a \emph{safety} (resp. a \emph{reachability}) game if all payoffs are $0$ or $-1$ (resp. $0$ or $1$).

\subsection{Strategies}

\begin{definition}[Strategies]
    A (history-dependent) \emph{strategy for Ada} (resp. Bertrand) in the game $\Game$ is a mapping $\sigma: \Hist \to \Dist(A)$ (resp. $\tau: \Hist \to \Dist(B)$).
    We call \emph{mixed action} for Ada (resp. Bertrand) an element of $\Dist(A)$ (resp. $\Dist(B)$).
    We refer to a pair $(\sigma, \tau)$ as a \emph{strategy profile}.
\end{definition}

A strategy $\upsilon$ (for Bertrand or for Ada) is \emph{pure} if for every history $h$, the support of the distribution $\upsilon(h)$ is a singleton.
It is \emph{memoryless} if for every two histories $h, h'$ with $\last(h) = \last(h')$, we have $\upsilon(h) = \upsilon(h')$. We identify memoryless strategies as a function over $S$. 
It is \emph{positional} if it is simultaneously pure and memoryless.
We denote the set of strategies for Ada by $\Strat_\Ada(\Game)$, and the set of strategies for Bertrand by $\Strat_\Bertrand(\Game)$ (or simply $\Strat_\Ada$ and $\Strat_\Bertrand$ when the context is clear).

We now define the mixture of two strategies.

\begin{definition}[Mixture of strategies] \label{def:statewise_mixture}
    Let $\sigma^0, \sigma^1$ be two memoryless strategies for Ada.
    Their \emph{statewise private mixture} is the strategy obtained by tossing a fair coin for every non-absorbing state at the beginning of the game.
    At every visit to a state $s$, it plays $\sigma^0(s)$ or $\sigma^1(s)$, as chosen by the coin of $s$.
    The coin outcomes remain private and are fixed throughout the play.
\end{definition}

The fact that this defines a strategy comes from Kuhn's theorem~\cite{Kuhn53}.

\subsection{Probabilities under a strategy profile}

A game $\Game$ and a strategy profile $(\sigma, \tau)$ define a probability measure $\prob_{\sigma\tau}$ over the set $\Plays(\Game)$.
For every random variable $X: \Plays(\Game) \to \RR$, we denote the expected value of $X$ under that probability measure by $\expec_{\sigma\tau}$.
We assimilate every history $h$ with the event of all plays that have $h$ as a prefix, and write $\prob_{\sigma\tau}(h)$ instead of $\prob_{\sigma\tau}(h (A \cdot B \cdot S)^\omega)$.
A history $h$ is \emph{compatible} with a strategy profile $(\sigma,\tau)$ if $\prob_{\sigma\tau}(h)>0$.

\subsection{Value}

As a fundamental property, recursive games are \emph{determined}: the largest payoff Ada can guarantee against every Bertrand strategy equals the smallest payoff Bertrand can guarantee against every Ada strategy.

\begin{theorem}
    In every recursive game $\Game$, we have:
    $\sup_\sigma \inf_\tau \expec_{\sigma\tau}(\mu) = \inf_\tau \sup_\sigma \expec_{\sigma\tau}(\mu).$
\end{theorem}

We call that quantity the \emph{value} of the game $\Game$.
In all what follows, we assume a recursive game $\Game$, and denote its value by $v$.

\subsection{Hyperoptimal strategies}

The classical solution concept in two-player zero-sum games is the notion of \emph{optimal} strategy.

\begin{definition}[Optimal strategy]
    An \emph{optimal} strategy for Ada is a strategy $\sigma$ such that for every Bertrand strategy $\tau$, we have $\expec_{\sigma\tau}(\mu) \geq v$.
\end{definition}

In this paper, we focus on a novel notion, stronger than that of optimal strategy.

\begin{definition}[Hyperoptimal strategy]
    A \emph{hyperoptimal} strategy for Ada is a strategy $\sigma$ such that for every Bertrand strategy $\tau$, we have $\expec_{\sigma\tau}(\mu) > v$.
\end{definition}

Those notions are defined analogously for Bertrand.
The following results are standard.

\begin{theorem}[\cite{DEALFARO2007188,DEALFARO2004374}]
    In negative recursive games, Ada always has an optimal strategy.
    By symmetry, in positive recursive games, Bertrand always has an optimal strategy.
    There exist negative recursive games where Ada has a hyperoptimal strategy.
\end{theorem}

From the introduction, we also know the following.

\begin{theorem}
    If Bertrand has an optimal strategy, then Ada has no hyperoptimal strategy.
    The converse is not true: there is an (absorbing and negative) recursive game in which Ada has no hyperoptimal strategy and Bertrand has no optimal strategy.
\end{theorem}


\section{Absorbing Games}\label{sec:absorbing}

In this section, we focus on absorbing recursive games.
We denote by $\Hist_{|s_0}$ the set of histories that remain in the state $s_0$, and by $\Hist^n_{|s_0}$ the set of such histories of length $n$.
Given a mixed action $x$ for Ada (resp. $y$ for Bertrand), we abuse notations and also denote by $x$ (resp. $y$) the memoryless strategy that plays $x$ (resp. $y$).

\subsection{Absorption rate, immediate reward, constraint}

Given a mixed action $x$ for Ada and an action $b$ for Bertrand, we define the \emph{absorption rate}:
$$\abs(x, b) = \sum_{s \neq s_0} \sum_{a \in A} x(a) \cdot \delta(s_0, a, b)(s),$$
and the \emph{immediate reward}:
$$\imm(x, b) = \sum_{s \neq s_0} \sum_{a \in A} x(a) \cdot \delta(s_0, a, b)(s) \cdot \mu(s).$$

Given a mixed action $y$ for Bertrand, we extend both notions by linearity:
$$\abs(x, y) = \sum_{b \in B} y(b) \cdot \abs(x, b),
\qquad
\imm(x, y) = \sum_{b \in B} y(b) \cdot \imm(x, b).$$
In particular, we have $\abs(x, \d_b) = \abs(x, b)$ and $\imm(x, \d_b) = \imm(x, b)$.
We now show how the expected payoff can be obtained from these two quantities.

\begin{lemma} \label{lm:imm_abs}
    If Ada plays the memoryless strategy $x$ and Bertrand plays the memoryless strategy $y$, then the expected payoff is $0$ if $\abs(x, y) = 0$, and otherwise:
    $$\expec_{x y}(\mu) = \frac{\imm(x, y)}{\abs(x, y)}.$$
\end{lemma}

\begin{proof}
    In the case $\abs(x, y) = 0$, the play never reaches an absorbing state, hence the expected payoff is $0$.
    If $\abs(x, y) > 0$, we have:
    $$\expec_{x y}(\mu) = \sum_{h \in \Hist_{|s_0}} \sum_{a \in A} \sum_{b \in B} \sum_{s \neq s_0} \prob_{x y}(h \, a \, b \, s) \cdot \mu(s)$$
    $$= \sum_{h \in \Hist_{|s_0}} \sum_{a \in A} \sum_{b \in B} \sum_{s \neq s_0} \prob_{x y}(h) \cdot x(a) \cdot y(b) \cdot \delta(s_0, a, b)(s) \cdot \mu(s)$$
    $$= \sum_{h \in \Hist_{|s_0}} \prob_{x y}(h) \cdot \imm(x, y).$$
    By grouping histories by length, we obtain:
    $$\expec_{x y}(\mu) = \imm(x, y) \cdot \sum_{n \geq 0} \sum_{h \in \Hist^n_{|s_0}} \prob_{x y}(h)$$
    $$= \imm(x, y) \cdot \sum_{n \geq 0} (1-\abs(x, y))^n$$
    $$= \frac{\imm(x, y)}{\abs(x, y)}$$
    since $\abs(x, y) > 0$.
\end{proof}


We now define the \emph{constraint} associated to the action $b$ as the mapping:
$$\cons_b: x \mapsto \imm(x, b) - v \cdot \abs(x, b),$$
and extend it to a mixed action $y$ for Bertrand by $\cons_y(x) = \imm(x, y) - v \cdot \abs(x, y)$, so that we have $\cons_y(x) = \sum_{b \in B} y(b) \cdot \cons_b(x)$.

Finally, a mixed action $x$ for Ada is \emph{positive} if it satisfies $\cons_b(x) \geq 0$ for every action $b$.
We will show later that optimal strategies are positive.
We write $P$ for the set of positive mixed actions, seen as a subset of $\RR^A$.
Note that each constraint $\cons_b$ is a linear mapping, as well as the equality $\sum_a x(a) = 1$, and therefore that the set $P \subseteq \RR^A$ is a polytope.

\begin{lemma} \label{lm:positive_nonempty}
    The polytope $P$ is nonempty.
\end{lemma}

\begin{proof}
    Consider the matrix game in which Ada selects an action $a \in A$, Bertrand selects an action $b \in B$, and Ada's payoff is $\cons_b(\d_a)$.
    Since $\cons$ is linear in the mixed action played by each player, the expected payoff obtained when Ada plays $x$ and Bertrand plays $y$ is $\cons_y(x)$.
    The polytope $P$ is therefore nonempty if and only if the value of that matrix game is non-negative.

    Let us assume towards a contradiction that it is negative, and let $y^*$ be an optimal mixed action for Bertrand in that matrix game: there is some $\epsilon > 0$ such that $\cons_{y^*}(x) \leq -\epsilon$ for every $x \in \Dist(A)$.
    We show that the memoryless strategy $y^*$ guarantees an expected payoff of at most $v - \epsilon$ against every strategy of Ada, which contradicts the fact that the value of the game is $v$.

    Since the strategy $y^*$ is memoryless, by optimality of positional strategies in Markov decision processes, memoryless strategies are sufficient for Ada to respond to it: we therefore only need to prove $\expec_{x y^*}(\mu) \leq v - \epsilon$ for every $x \in \Dist(A)$.
    Let therefore $x \in \Dist(A)$.
    Every mixed action $x$ satisfying $\abs(x, y^*) = 0$ also satisfies $\imm(x, y^*) = 0$, and therefore $\cons_{y^*}(x) = 0$: since we have $\cons_{y^*}(x) \leq -\epsilon$, we obtain $\abs(x, y^*) > 0$.
    By \Cref{lm:imm_abs}, we then have:
    \[
        \expec_{x y^*}(\mu)
        = \frac{\imm(x, y^*)}{\abs(x, y^*)}
        \leq \frac{v \cdot \abs(x, y^*) - \epsilon}{\abs(x, y^*)}
        = v - \frac{\epsilon}{\abs(x, y^*)}
        \leq v - \epsilon,
    \]
    where the first inequality follows from $\cons_{y^*}(x) \leq -\epsilon$, and the second one from $\abs(x, y^*) \leq 1$.
\end{proof}

We now establish the link between positivity and optimality.

\begin{lemma} \label{lm:positive_strategies}
    Let $x \in \Dist(A)$.
    If the memoryless strategy $x$ is optimal, then the mixed action $x$ is positive.
    If we have $v < 0$, then the converse holds.
\end{lemma}

\begin{proof}
    We first prove the statement, then the converse with $v < 0$.

    \subp{If the memoryless strategy $x$ is optimal}
    Let $b$ be an action for Bertrand, and let us show $\cons_b(x) \geq 0$.
    Let us first rule out the case $\abs(x, b) = 0$: then, we also have $\imm(x, b) = 0$, and therefore $\cons_b(x) = 0$.
    We now assume $\abs(x, b) > 0$.
    Then, by \Cref{lm:imm_abs}, we have:
    $$\expec_{x\d_b}(\mu) = \frac{\imm(x, b)}{\abs(x, b)}.$$
    Since the memoryless strategy $x$ is optimal, it guarantees at least the expected payoff $v$ against the positional strategy $\d_b$, hence the quantity above is at least $v$.
    Hence $\cons_b(x) \geq 0$.

    \subp{If we have $v < 0$ and the mixed action $x$ is positive}

    Since the strategy $x$ is memoryless, by optimality of positional strategies in Markov decision processes, positional strategies are sufficient to respond to $x$: we therefore only need to prove $\expec_{x\d_b}(\mu) \geq v$ for every Bertrand action $b$.
    Let therefore $b \in B$.
    
    If we have $\abs(x, b) = 0$, then when following the strategy profile $(x, b)$, no terminal state is ever reached, and the payoff obtained is $0$, which is larger than $v$.
    Let us now assume $\abs(x, b) > 0$.
    Then, by \Cref{lm:imm_abs} again, we have:
    $$\expec_{xb}(\mu) = \frac{\imm(x, b)}{\abs(x, b)}.$$
    Since we have $\cons_b(x) \geq 0$, we have $\imm(x, b) \geq v \cdot \abs(x, b)$, and the above quantity is greater than or equal to $v$.
\end{proof}

\subsection{Critical actions}

We now consider the actions for Bertrand that Ada has to consider in priority.
We define the \emph{critical actions} of Bertrand as the actions $b \in B$ such that $\cons_b(p) = 0$ for every $p \in P$.
The set of critical actions is written $B_\crit$.

\begin{lemma} \label{lm:critical_relint}
    Let $p^\circ$ belong to the relative interior of $P$.
    For every action $b\in B$, we have $b\in B_\crit$ if and only if $\cons_b(p^\circ)=0$.
\end{lemma}

\begin{proof}
    The direct implication follows from the definition of critical actions.
    Conversely, assume that $\cons_b(p^\circ)=0$ and let $p\in P$.
    Since $p^\circ$ belongs to the relative interior of $P$, for every sufficiently small $\lambda>0$, the point
    $p_\lambda=p^\circ+\lambda(p^\circ-p)$ belongs to $P$.
    By linearity of $\cons_b$, we have
    \begin{align*}
        0
        &\leq \cons_b(p_\lambda) \\
        &=(1+\lambda)\cons_b(p^\circ)-\lambda\cons_b(p) \\
        &=-\lambda\cons_b(p),
    \end{align*}
    and therefore $\cons_b(p) \leq 0$.
    Since $p\in P$, we also have $\cons_b(p)\geq0$, and therefore $\cons_b(p)=0$.
    Thus, $\cons_b = 0$ on every element of $P$, and hence $b\in B_\crit$.
\end{proof}

Importantly, Bertrand always have critical actions.

\begin{lemma} \label{lm:critical_nonempty}
    The set $B_\crit$ is nonempty.
\end{lemma}

\begin{proof}
    By \Cref{lm:positive_nonempty}, the polytope $P$ is nonempty: let $p^\circ$ belong to its relative interior.
    By optimality of positional strategies in MDPS, the memoryless strategy $p^\circ$ has a positional best response: let us denote it $\d_b$.

    Let us first prove that we have $\cons_b(p^\circ) = 0$.
    Since $p^\circ$ is positive, we have $\cons_b(p^\circ) \geq 0$: let us assume $\cons_b(p^\circ) > 0$.
    Then, we have $\abs(p^\circ, b) > 0$, and by \Cref{lm:imm_abs}:
    $$\expec_{p^\circ b}(\mu) = \frac{\imm(p^\circ, b)}{\abs(p^\circ, b)}.$$
    Since the positional strategy $\d_b$ is a best response to $p^\circ$, this quantity is bounded by the value of the game:
    $$\frac{\imm(p^\circ, b)}{\abs(p^\circ, b)} \leq v,$$
    i.e. $\imm(p^\circ, b) \leq v \cdot \abs(p^\circ, b)$ and therefore $\cons_b(p^\circ) \leq 0$, which is a contradiction.

    By \Cref{lm:critical_relint}, the equality $\cons_b(p^\circ)=0$ implies that $b\in B_\crit$.
\end{proof}

\subsection{The case $v \geq 0$}

We can now rule out the case where the value is non-negative.

\begin{theorem} \label{thm:abs_v_nonneg}
    If we have $v \geq 0$, then Ada has no hyperoptimal strategy.
\end{theorem}

\begin{proof}
    Let $\sigma$ be a strategy for Ada.
    We construct a strategy $\tau$ for Bertrand such that
    $\expec_{\sigma\tau}(\mu) \leq v$.
    For every history $h \in \Hist_{|s_0}$, let us choose an action $b_h$ such that
    $$\cons_{b_h}(\sigma(h)) \leq 0.$$
    Such an action always exists.
    Indeed, if the mixed action $\sigma(h)$ is not positive, this follows from the definition of positivity.
    Otherwise, we choose $b_h \in B_\crit$, which is possible by \Cref{lm:critical_nonempty}, and then we have $\cons_{b_h}(\sigma(h)) = 0$ by definition of critical actions.
    We define $\tau(h) = \d_{b_h}$ for every $h \in \Hist_{|s_0}$, and define $\tau$ arbitrarily on all other histories.

    Then, we have that
    \begin{align*}
        \expec_{\sigma\tau}(\mu)
        &= \sum_{h \in \Hist_{|s_0}} \sum_{a \in A} \sum_{s \neq s_0}
        \prob_{\sigma\tau}(h a b_h s) \cdot \mu(s) \\
        &= \sum_{h \in \Hist_{|s_0}} \prob_{\sigma\tau}(h)
        \sum_{a \in A} \sum_{s \neq s_0} \sigma(h)(a) \cdot
        \delta(s_0, a, b_h)(s) \cdot \mu(s) \\
        &= \sum_{h \in \Hist_{|s_0}} \prob_{\sigma\tau}(h)
        \cdot \imm(\sigma(h), b_h) \\
        &\leq v \cdot \sum_{h \in \Hist_{|s_0}} \prob_{\sigma\tau}(h)
        \cdot \abs(\sigma(h), b_h) \\
        &\leq v,
    \end{align*}
    where the first inequality follows from the choice of $b_h$. The second inequality follows
    from (i)~the assumpion $v \geq 0$; and (ii)~the sum is the probability that an absorbing state is eventually reached, and is therefore at most $1$.
    Hence, every strategy of Ada admits a response achieving an expected payoff at most $v$, so Ada has no hyperoptimal strategy.
\end{proof}

In what follows, we focus on the negative case.

\subsection{The case $v < 0$}

We now assume $v < 0$.
Note that by \Cref{lm:positive_strategies}, the polytope $P$ is now the set of optimal memoryless strategies.
Optimal strategies of particular interest are those that lie in the relative interior of $P$, since their support is maximal among positive strategies.

\begin{lemma} \label{lm:relint_best_response}
    Let $p^\circ$ belong to the relative interior of $P$.
    Every best response $\tau$ to $p^\circ$ plays only critical actions along histories compatible with $(p^\circ,\tau)$.
\end{lemma}

\begin{proof}
    Since $v<0$ and $p^\circ\in P$, the memoryless strategy $p^\circ$ is optimal by \Cref{lm:positive_strategies}.
    Hence, if $\tau$ is a best response to $p^\circ$, then $\expec_{p^\circ\tau}(\mu)=v$.
    Let $\zeta$ be the probability that an absorbing state is eventually reached, when following the strategy profile $(p^\circ,\tau)$. 
    Decomposing the expected payoff according to the history $h \in \Hist_{|s_0}$ after which absorption occurs, and using the fact that $p^\circ$ is memoryless, we obtain
    \begin{align*}
        \expec_{p^\circ\tau}(\mu)
        &= \sum_{h \in \Hist_{|s_0}} \sum_{a \in A} \sum_{b \in B} \sum_{s \neq s_0}
        \prob_{p^\circ\tau}(h) \cdot p^\circ(a) \cdot \tau(h)(b) \cdot \delta(s_0, a, b)(s) \cdot \mu(s) \\
        &= \sum_{h \in \Hist_{|s_0}} \prob_{p^\circ\tau}(h) \cdot \imm(p^\circ, \tau(h)),
    \end{align*}
    the last equality being the definition of the immediate reward.
    The same decomposition, applied to the event that an absorbing state is eventually reached rather than to the payoff, gives
    \[
        \zeta = \sum_{h \in \Hist_{|s_0}} \prob_{p^\circ\tau}(h) \cdot \abs(p^\circ, \tau(h)).
    \]
    Now, by definition of the constraint, every history $h \in \Hist_{|s_0}$ satisfies
    $$\imm(p^\circ, \tau(h)) = \cons_{\tau(h)}(p^\circ) + v \cdot \abs(p^\circ, \tau(h)).$$
    Substituting this identity in the first equation, and then using the second one, we obtain
    \[
        \expec_{p^\circ\tau}(\mu)
        =v \cdot \zeta +\sum_{h\in\Hist_{|s_0}}\prob_{p^\circ\tau}(h) \cdot\cons_{\tau(h)}(p^\circ).
    \]
    The terms in the sum are non-negative because $p^\circ\in P$ and for every history $h \in \Hist_{|s_0}$ we have $\cons_{\tau(h)}(p^\circ) \geq 0$, and $v \cdot \zeta\geq v$
    because $v<0$ and $\zeta\leq1$.
    Moreover, by \Cref{lm:critical_relint}, we have $\cons_b(p^\circ)>0$ for every non-critical action $b$.
    Thus, if $\tau$ played a non-critical action with positive probability after a compatible history, the displayed expectation would be strictly larger than $v$, a contradiction.
\end{proof}

We can now define another subclass of optimal strategies: \emph{stalling} strategies, designed to force a loop as soon as Bertrand restricts himself to critical actions.

\begin{definition}[Stalling action, stalling strategy]
    A \emph{stalling action} is an action $a$ for Ada such that for every critical action $b \in B_\crit$, we have $\delta(s_0, a, b)(s_0) = 1$.
    We denote the set of stalling actions by $A_\stall$.
    A \emph{stalling strategy} is a memoryless strategy $p \in P$ such that $\Supp(p) \subseteq A_\stall$.
\end{definition}

Finally, we prove our characterisation of hyperoptimality in absorbing games, by establishing the link between stalling strategies and hyperoptimal strategies.

\begin{theorem}
    Ada has a hyperoptimal strategy if and only if she has a stalling strategy.
\end{theorem}

\begin{proof}
    We first show how to construct a hyperoptimal strategy from a stalling strategy.
    Then, we show how the existence of the former implies the existence of the latter.

    \subp{From stalling to hyperoptimal}
    Let $p$ be a stalling strategy, and choose $p^\circ$ in the relative interior of $P$.
    Let $\sigma$ be the fair mixture of the memoryless strategies $p$ and $p^\circ$.
    For every strategy $\tau$ of Bertrand,
    \begin{equation} \label{eq:initial_mixture}
        \expec_{\sigma\tau}(\mu)
        =\frac12\expec_{p\tau}(\mu)
         +\frac12\expec_{p^\circ\tau}(\mu).
    \end{equation}
    Both $p$ and $p^\circ$ are optimal by \Cref{lm:positive_strategies}, so $\sigma$ is optimal.

    Assume towards a contradiction that $\sigma$ has a best response $\tau$.
    Then $\expec_{\sigma\tau}(\mu)=v$.
    Since both terms on the right-hand side of \Cref{eq:initial_mixture} are at least $v$, they must both be equal to $v$.
    Thus, $\tau$ is a best response to both $p$ and $p^\circ$.

    Then, by \Cref{lm:relint_best_response}, the strategy $\tau$ plays only critical actions along histories compatible with $(p^\circ,\tau)$.
    Since $\Supp(p)\subseteq\Supp(p^\circ)$, this holds in particular along histories compatible with $(p,\tau)$.
    Since $p$ is stalling, no absorbing state is reached with positive probability under $(p,\tau)$, and hence $\expec_{p\tau}(\mu)=0$.
    This contradicts the fact that $\tau$ is a best response to $p$, since $v<0$.

    Thus, the optimal strategy $\sigma$ has no best response, and is therefore hyperoptimal.

    \subp{From hyperoptimal to stalling}
    Let $\sigma$ be a hyperoptimal strategy, and let $y$ be the uniform distribution over the set $B_\crit$, which is nonempty by \Cref{lm:critical_nonempty}.
    Since the state $s_0$ is the only non-absorbing state, for every history $h \in \Hist_{|s_0}$, the continuation of $\sigma$ after $h$ is again a strategy of the game $\Game$: let us denote it by $\sigma_h$, and let us define
    $$w(h) = \inf_\tau \expec_{\sigma_h \tau}(\mu).$$
    By definition of the value, we have that $w(h) \leq v$ for every history $h \in \Hist_{|s_0}$.
    Moreover, since the strategy $\sigma$ is optimal, we have that $w(s_0) = v$.

    \begin{claim} \label{clm:critical_continuations}
        For every history $h \in \Hist_{|s_0}$ compatible with $(\sigma, y)$, we have $w(h) = v$ and $\sigma(h) \in P$.
    \end{claim}

    \begin{claimproof}
        We prove both assertions simultaneously by induction on the length of the histories compatible with $(\sigma, y)$.

        Let $h \in \Hist_{|s_0}$ be any history, and let us write $x = \sigma(h)$.
        Fix an action $b$ for Bertrand.
        Let $\epsilon > 0$, and consider a response to $\sigma_h$ that starts by playing $b$.
        After every non-absorbing successor history $h a b s_0$, Bertrand can follow a response to $\sigma_{h a b s_0}$ whose expected payoff is at most $w(h a b s_0) + \epsilon$.
        Such responses exist by definition of these quantities as infima.
        Therefore, we have 
        $$w(h) \leq \imm(x, b) + \sum_{a \in A} x(a) \cdot \delta(s_0, a, b)(s_0) \cdot \bigl(w(h a b s_0) + \epsilon\bigr).$$
        Letting $\epsilon$ tend to $0$, we obtain
        \begin{equation} \label{eq:critical_continuation_bound}
            \begin{aligned}
                w(h)
                &\leq \imm(x, b) + \sum_{a \in A} x(a) \cdot \delta(s_0, a, b)(s_0) \cdot w(h a b s_0) \\
                &\leq \imm(x, b) + \bigl(1 - \abs(x, b)\bigr) \cdot v \\
                &= v + \cons_b(x),
            \end{aligned}
        \end{equation}
        where the second inequality follows from $w(h a b s_0) \leq v$ for every action $a \in A$ (by definition of the value), and from the fact that the coefficients in the sum add up to the probability $1 - \abs(x, b)$ of remaining in $s_0$.

        For the initial history $s_0$, we have $w(s_0) = v$ because $\sigma$ is optimal.
        Applying \Cref{eq:critical_continuation_bound} at $s_0$ gives
        $$v = w(s_0) \leq v + \cons_b(\sigma(s_0))$$
        for every action $b$.
        Subtracting $v$ gives $\cons_b(\sigma(s_0)) \geq 0$ for every $b$, hence $\sigma(s_0) \in P$.
        Both assertions therefore hold at length $1$.

        Assume now that both assertions hold for every compatible history of length $n$, and let $h' = h a b s_0$ be a compatible history of length $n+1$.
        Its prefix $h$ is compatible and has length $n$, so the induction hypothesis gives $w(h) = v$ and $x = \sigma(h) \in P$.
        Moreover, compatibility of $h'$ gives $x(a) \cdot \delta(s_0, a, b)(s_0) > 0$ and $y(b) > 0$.
        Since $y$ plays only critical actions, we have $b \in B_\crit$, and therefore $\cons_b(x) = 0$.
        Thus, the chain of inequalities in \Cref{eq:critical_continuation_bound}, applied at $h$ with the action $b$, starts and ends at $v$, so both inequalities are equalities.
        In particular, replacing the continuation values $w(h a b s_0)$ by $v$ does not increase the sum, which gives
        $$\sum_{a' \in A} x(a') \cdot \delta(s_0, a', b)(s_0) \cdot \bigl(v - w(h a' b s_0)\bigr) = 0.$$
        Every term is non-negative because $w(h a' b s_0) \leq v$.
        The coefficient of the term with $a' = a$ is positive, so $w(h') = v$.

        It remains to show that $\sigma(h') \in P$.
        Applying \Cref{eq:critical_continuation_bound} at $h'$ gives, for every action $b'$,
        $$v = w(h') \leq v + \cons_{b'}(\sigma(h')).$$
        Hence, $\cons_{b'}(\sigma(h')) \geq 0$ for every $b'$, and therefore $\sigma(h') \in P$.
        Both assertions hold at $h'$, completing the induction.
    \end{claimproof}

    Let us now show that every $p \in P$ satisfying $\abs(p, y) = 0$ is a stalling strategy.
    Let $p \in P$ be such a mixed action.
    Since the mixed action $y$ gives a positive probability to every critical action, the equality $\abs(p, y) = 0$ gives $\abs(p, b) = 0$ for every action $b \in B_\crit$.
    Hence, we have $\delta(s_0, a, b)(s_0) = 1$ for every action $a \in \Supp(p)$ and every action $b \in B_\crit$, i.e. $\Supp(p) \subseteq A_\stall$ which implies that the memoryless strategy $p$ is stalling.

    Let now $h \in \Hist_{|s_0}$ be a history compatible with $(\sigma, y)$.
    By \Cref{clm:critical_continuations}, we have $\sigma(h) \in P$, and therefore $\cons_y(\sigma(h)) = 0$ by definition of critical actions.
    Let us write $\zeta$ for the probability that an absorbing state is eventually reached, when following the strategy profile $(\sigma, y)$: let us show that $\zeta < 1$.
    We then have that
    \begin{align*}
        \expec_{\sigma y}(\mu)
        &= \sum_{h \in \Hist_{|s_0}} \prob_{\sigma y}(h) \cdot \imm(\sigma(h), y) \\
        &= v \cdot \sum_{h \in \Hist_{|s_0}} \prob_{\sigma y}(h) \cdot \abs(\sigma(h), y) \\
        &= v \cdot \zeta,
    \end{align*}
    where the second equality follows from the equality $\cons_y(\sigma(h)) = 0$, which holds for every history $h$ compatible with $(\sigma, y)$, the other terms of the sum being zero.
    Since the strategy $\sigma$ is hyperoptimal, we have $\expec_{\sigma y}(\mu) > v$, and therefore $v \cdot \zeta > v$. Since $v < 0$, we obtain $\zeta < 1$.

    Finally, the mapping $\abs(\cdot, y)$ is linear, hence continuous, and the polytope $P$ is compact and nonempty by \Cref{lm:positive_nonempty}. Therefore, the quantity
    $$\eta = \min_{p \in P} \abs(p, y)$$
    is well defined, and it is attained by some $p^\star \in P$.
    After every history compatible with $(\sigma, y)$, the conditional probability that an absorbing state is reached at the next round is at least $\eta$. Therefore, the probability that no absorbing state is reached during the first $n$ rounds is at most $(1-\eta)^n$ for every integer $n$.
    That probability is also at least $1-\zeta$, since with probability $1-\zeta$ no absorbing state is ever reached.
    We thus have $(1-\eta)^n \geq 1-\zeta > 0$ for every integer $n$, which forces $\eta = 0$.
    Consequently, we have $\abs(p^\star, y) = 0$, and by the above, the memoryless strategy $p^\star$ is a stalling strategy.
\end{proof}






\section{Negative Recursive Games}\label{sec:negative}
In this section, we consider negative recursive games.
For every state $s$, we denote by $\Game_s$ the game obtained from $\Game$ by replacing the initial state with $s$.
We denote the value of that game by $v_s$.
We write $S_\Abs$ for the set of absorbing states, and use the convention $v_s = \mu(s)$ for every $s \in S_\Abs$.
In particular, we have $v_s \leq 0$ for every state $s$, and $v_{s_0} = v$.
When such an initial state $s \neq s_0$ is specified, strategies and expectations are understood in the game $\Game_s$.

We show that whenever Ada has a hyperoptimal strategy, she has one obtained by a mixture of two optimal memoryless strategies, as in \Cref{def:statewise_mixture}.
Thus, two mixed actions are fixed at each non-absorbing state, and an independent private fair coin chooses the mixed action used at that state throughout the play.
The first mixed action belongs to the relative interior of the polytope of locally optimal mixed actions.
The second is obtained from a finite auxiliary game whose Ada actions are the vertices of these polytopes and whose Bertrand actions are critical.

\subsection{Constraints and positive mixed actions}

Given a state $s$, a mixed action $x$ for Ada and an action $b$ for Bertrand, we extend the transition function by linearity:
$$\delta(s, x, b)(t) = \sum_{a \in A} x(a) \cdot \delta(s, a, b)(t).$$
We define the \emph{constraint} associated to $s$ and $b$ as the mapping:
$$\cons_{s,b}: x \mapsto \sum_{a \in A} x(a) \cdot \left(\sum_{t \in S} \delta(s, a, b)(t) \cdot v_t - v_s\right),$$
and extend it to a mixed action $y$ for Bertrand by $\cons_{s,y}(x) = \sum_{b \in B} y(b) \cdot \cons_{s,b}(x)$.
Thus, for every mixed action $x$, we have:
$$\sum_{t \in S} \delta(s, x, b)(t) \cdot v_t = v_s + \cons_{s,b}(x).$$

A mixed action $x$ for Ada is \emph{positive} at $s$ if it satisfies $\cons_{s,b}(x) \geq 0$ for every action $b$.
We denote the polytope of positive mixed actions at $s$ by $P_s$.
A strategy $\sigma$ for Ada is \emph{locally optimal} if $\sigma(h) \in P_{\last(h)}$ for every history $h$.
Note that at an absorbing state $s$, all constraints vanish and we have $P_s = \Dist(A)$.

\begin{lemma} \label{lm:negative_positive_nonempty}
    For every state $s$, the matrix game with payoff
    $$M_s(a,b) = \sum_{t \in S} \delta(s,a,b)(t) \cdot v_t$$
    has value $v_s$.
    Consequently, the polytope $P_s$ is nonempty and consists exactly of Ada's optimal mixed actions in that matrix game.
\end{lemma}

\begin{proof}
    Let $m_s$ be the value of the matrix game, and let $\epsilon > 0$.
    Ada can play an optimal mixed action of that matrix game at the first round and, after reaching a state $t$, follow an $\epsilon$-optimal strategy in $\Game_t$.
    This guarantees an expected payoff of at least $m_s - \epsilon$, so $v_s \geq m_s - \epsilon$.
    Symmetrically, Bertrand can play an optimal mixed action of the matrix game and then an $\epsilon$-optimal strategy from the state reached, giving $v_s \leq m_s + \epsilon$.
    Letting $\epsilon$ tend to $0$, we obtain $m_s = v_s$.
    The final assertion follows from the definition of the constraints and the existence of optimal mixed actions in finite matrix games.
\end{proof}

We now prove the following standard result, under the form that we will use.

\begin{lemma} \label{lm:negative_positive_optimal}
    Every locally optimal strategy for Ada is optimal from every state.
    In particular, Ada has an optimal memoryless strategy from every state.
\end{lemma}

\begin{proof}
    Let $\sigma$ be a locally optimal strategy, let $\tau$ be a strategy for Bertrand, and fix an initial state $s$.
    Let $S_n$ and $H_n$ denote the state and history at round $n$, and let $V_n = v_{S_n}$.
    By the definition of the constraints, we have:
    $$\expec_{\sigma\tau}\bigl(V_{n+1} \mid H_n\bigr)
    = V_n + \cons_{S_n,\tau(H_n)}\bigl(\sigma(H_n)\bigr) \geq V_n.$$
    Hence, the sequence $\expec_{\sigma\tau}(V_n)$ is non-decreasing and is bounded below by $v_s$.

    Let $\mu_n$ be the payoff of the absorbing state reached by round $n$, if there is one, and $0$ otherwise.
    Since all values are non-positive, we have $\mu_n \geq V_n$ on every play.
    Moreover, $\mu_n$ converges to $\mu$, and these random variables are uniformly bounded because the set of absorbing rewards is finite.
    By dominated convergence, we obtain:
    $$\expec_{\sigma\tau}(\mu)
    = \lim_{n \to \infty} \expec_{\sigma\tau}(\mu_n)
    \geq v_s.$$
    Thus, $\sigma$ is optimal from $s$.
    Finally, choosing one mixed action from each nonempty polytope $P_s$ gives a locally optimal memoryless strategy.
\end{proof}

\subsection{Critical actions and interior strategies}

We define the \emph{critical actions at $s$} as the actions $b$ for Bertrand such that $\cons_{s,b}(p) = 0$ for every $p \in P_s$, and denote their set by $B^s_\crit$.
A strategy $\tau$ for Bertrand is \emph{critical} if $\Supp(\tau(h)) \subseteq B^{\last(h)}_\crit$ at every non-absorbing history $h$.

\begin{lemma} \label{lm:negative_critical_relint}
    Let $p^\circ_s$ belong to the relative interior of $P_s$.
    For every action $b$, we have $b \in B^s_\crit$ if and only if $\cons_{s,b}(p^\circ_s) = 0$.
    Moreover, $B^s_\crit$ is nonempty, and $\Supp(p) \subseteq \Supp(p^\circ_s)$ for every $p \in P_s$.
\end{lemma}

\begin{proof}
    The equivalence follows from the same argument as \Cref{lm:critical_relint}.
    Indeed, for every $p \in P_s$ and every sufficiently small $\lambda > 0$, the mixed action
    $$p_\lambda = p^\circ_s + \lambda(p^\circ_s - p)$$
    belongs to $P_s$.
    If $\cons_{s,b}(p^\circ_s) = 0$, linearity gives
    $$0 \leq \cons_{s,b}(p_\lambda) = -\lambda \cdot \cons_{s,b}(p),$$
    and positivity of $p$ then gives $\cons_{s,b}(p) = 0$.
    The other implication follows from the definition of critical actions.

    By \Cref{lm:negative_positive_nonempty}, the matrix game $M_s$ has value $v_s$.
    Consider now the matrix game in which Ada selects an action $a \in A$, Bertrand selects an action $b \in B$, and Ada's payoff is $\cons_{s,b}(\d_a) = M_s(a, b) - v_s$. The value of this matrix game is therefore $0$.
    Since $p^\circ_s \in P_s$, we have $\cons_{s,b}(p^\circ_s) \geq 0$ for every action $b$.
    Moreover, since $\cons_s$ is linear in the mixed action played by each player, the value of that matrix game is $\max_x \min_{b \in B} \cons_{s,b}(x)$, and is therefore at least $\min_{b \in B} \cons_{s,b}(p^\circ_s)$.
    Hence, we have $\min_{b \in B} \cons_{s,b}(p^\circ_s) = 0$, and by the equivalence proved above, every action $b$ attaining that minimum belongs to $B^s_\crit$, which is therefore nonempty.

    Finally, if $p(a) > 0$ and $p^\circ_s(a) = 0$ for some $p \in P_s$, then the mixed action $p_\lambda$ above satisfies $p_\lambda(a) = -\lambda \cdot p(a) < 0$, which is a contradiction.
    Thus, $\Supp(p) \subseteq \Supp(p^\circ_s)$.
\end{proof}

In what follows, we fix, at every state $s$, a mixed action $p^\circ_s$ in the relative interior of $P_s$.
We denote by $p^\circ: s \mapsto p^\circ_s$ the associated memoryless strategy.
It is optimal by \Cref{lm:negative_positive_optimal}.

\begin{lemma} \label{lm:negative_relint_best_response}
    Every best response $\tau$ to $p^\circ$ plays only critical actions along non-absorbing histories compatible with $(p^\circ,\tau)$.
\end{lemma}

\begin{proof}
    Since $p^\circ$ is optimal, a best response $\tau$ satisfies $\expec_{p^\circ\tau}(\mu) = v_{s_0}$.
    With the notation of the proof of \Cref{lm:negative_positive_optimal}, for every $n$, we have:
    $$\expec_{p^\circ\tau}(V_n)
    = v_{s_0} + \sum_{k=0}^{n-1} \expec_{p^\circ\tau}\left(\cons_{S_k,\tau(H_k)}(p^\circ(S_k))\right).$$
    All terms in that sum are non-negative.
    Moreover, for every $m \geq n$, we have $\expec_{p^\circ\tau}(\mu_m) \geq \expec_{p^\circ\tau}(V_m) \geq \expec_{p^\circ\tau}(V_n)$.
    Letting $m$ tend to infinity gives $\expec_{p^\circ\tau}(\mu) \geq \expec_{p^\circ\tau}(V_n)$.
    Thus, every term in the displayed sum is zero.

    Let $h$ be a non-absorbing history compatible with $(p^\circ, \tau)$, and let $s = \last(h)$.
    We then have $\cons_{s,\tau(h)}(p^\circ(s)) = 0$.
    By \Cref{lm:negative_critical_relint}, every non-critical action $b$ satisfies $\cons_{s,b}(p^\circ(s)) > 0$, so $\tau(h)$ gives probability $0$ to every such action.
\end{proof}

\subsection{Optimal continuations and strict improvement}

In what follows, we will show how to define an optimal memoryless strategy $p$ that can be mixed with $p^\circ$, when it exists, to obtain a hyperoptimal strategy.
The core idea remains the same as in the absorbing case: since Ada may play $p^\circ$, Bertrand is forced to play critical actions, and the strategy $p$ will take advantage of that fact to trap the game into some region.

Let $T = S_\Abs \cup \{s \in S \mid v_s = 0\}$,
and let $n_T$ be the first round at which a state of $T$ is reached, with $n_T = \infty$ if no such state is reached.
The inclusion of zero-value states in $T$ is essential because, from such a state, a locally optimal strategy gives payoff $0$ against every response, so no strict improvement over its value is possible.

\begin{lemma} \label{lm:negative_strict_improvement}
    Assume $S \setminus T \neq \emptyset$, and let $\eta = \min_{s \notin T}(-v_s) > 0$.
    For every locally optimal strategy $\sigma$ and every strategy $\tau$, we have:
    $$\expec_{\sigma\tau}(\mu) \geq v_{s_0} + \eta \cdot \prob_{\sigma\tau}(n_T = \infty).$$
\end{lemma}

\begin{proof}
    Use the random variables $V_n$ and $\mu_n$ defined in the proof of \Cref{lm:negative_positive_optimal}.
    On the event $n_T > n$, the current state is non-absorbing and has value at most $-\eta$, so $\mu_n - V_n \geq \eta$.
    On all other plays, we still have $\mu_n - V_n \geq 0$.
    Since $\expec_{\sigma\tau}(V_n) \geq v_{s_0}$, we obtain:
    $$\expec_{\sigma\tau}(\mu_n) \geq v_{s_0} + \eta \cdot \prob_{\sigma\tau}(n_T > n).$$
    Letting $n$ tend to infinity proves the statement by dominated convergence.
\end{proof}

\begin{lemma} \label{lm:negative_critical_continuations}
    Let $\sigma$ be optimal and let $\tau$ be critical.
    After every history $h$ compatible with $(\sigma,\tau)$, the continuation of $\sigma$ is optimal from $\last(h)$, and $\sigma(h) \in P_{\last(h)}$.
\end{lemma}

\begin{proof}
    For every history $h$, let $\sigma_h$ denote its continuation, and write
    $$w(h) = \inf_{\tau'} \expec_{\sigma_h\tau'}(\mu).$$
    We have $w(h) \leq v_{\last(h)}$ for every history, and $w(s_0) = v_{s_0}$.
    Let $s = \last(h)$, let $x = \sigma(h)$, and fix an action $b$.
    Let $\epsilon > 0$.
    Bertrand can play $b$ and then, after every successor history $h a b t$, choose a response whose expected payoff is at most $w(h a b t) + \epsilon$, which exists by definition of the infimum.
    Letting $\epsilon$ tend to $0$, we obtain:
    \begin{align*}
        w(h)
        &\leq \sum_{a \in A}\sum_{t \in S} x(a) \cdot \delta(s,a,b)(t) \cdot w(h a b t) \\
        &\leq \sum_{a \in A}\sum_{t \in S} x(a) \cdot \delta(s,a,b)(t) \cdot v_t \\
        &= v_s + \cons_{s,b}(x).
    \end{align*}
    If $w(h) = v_s$, this gives $x \in P_s$.
    Moreover, if $b \in B^s_\crit$, then both inequalities are equalities, and therefore
    $$\sum_{a \in A}\sum_{t \in S} x(a) \cdot \delta(s,a,b)(t) \cdot \bigl(v_t-w(h a b t)\bigr) = 0.$$
    Every term is non-negative, so $w(h a b t) = v_t$ whenever $x(a) \cdot \delta(s,a,b)(t) > 0$.
    Starting with $w(s_0) = v_{s_0}$, an induction on the length of compatible histories proves the statement, since $\tau$ plays only critical actions.
    At absorbing states, the continuation payoff is fixed and the assertion is immediate.
\end{proof}

\subsection{The auxiliary game}

We now define a variant of the game $\Game$, where Ada is constrained to play a locally optimal strategy, and Bertrand to play critical actions.

For every state $s \notin T$, let $E_s$ be the finite set of vertices of $P_s$.
We define the \emph{auxiliary game} $\bar\Game$ on the same states as $\Game$, making all states of $T$ absorbing.
At a state $s \notin T$, Ada's actions are the elements of $E_s$, Bertrand's actions are the elements of $B^s_\crit$, and the transition function is:
$$\bar\delta(s,p,b)(t) = \delta(s,p,b)(t)
= \sum_{a \in A} p(a) \cdot \delta(s,a,b)(t).$$
Bertrand's objective in this game is to reach $T$ almost surely.
Let $W$ be his almost-sure winning set: a state belongs to $W$ if Bertrand has a strategy that reaches $T$ with probability $1$ against every strategy of Ada.
By the qualitative theory of concurrent reachability games~\cite[Theorem~8]{DEALFARO2007188}, Bertrand has a memoryless winning strategy on $W$, and Ada has a strategy that avoids $T$ forever with positive probability from every state outside $W$ against every strategy of Bertrand.
The latter probability may depend on Bertrand's strategy.

\begin{lemma} \label{lm:negative_reaching_target}
    From every state in $W$, Bertrand has a memoryless critical strategy that reaches $T$ almost surely whenever Ada plays positive mixed actions along compatible histories.
\end{lemma}

\begin{proof}
    Let $y$ be a memoryless strategy for Bertrand that wins in $\bar\Game$ from every state of $W$, and let us extend it to the non-absorbing states of $\Game$ that belong to $T$ by arbitrary critical mixed actions, which is possible by \Cref{lm:negative_critical_relint}.
    The memoryless strategy $y$ is then critical in $\Game$.
    Let $\sigma$ be a strategy for Ada that plays positive mixed actions along the histories compatible with $(\sigma, y)$: let us show that when following $(\sigma, y)$, the set $T$ is reached almost surely.
    We construct a strategy $\bar\sigma$ for Ada in $\bar\Game$ that induces the same distribution over state sequences as $\sigma$, up to the first visit to $T$.

    Let $\pi$ be a finite state sequence that has positive probability under $(\sigma, y)$ and visits no state of $T$, and let $s$ be its last state.
    We denote by $x_\pi$ the average of the mixed actions $\sigma(h)$ over the histories $h$ compatible with $(\sigma, y)$ whose state sequence is $\pi$, each of them weighted by its probability conditional on $\pi$.
    Every such history ends at $s$, so we have $\sigma(h) \in P_s$ by the hypothesis on $\sigma$: the polytope $P_s$ being convex, we obtain $x_\pi \in P_s$.
    The mixed action $x_\pi$ is therefore a convex combination of the vertices of $P_s$, and we let $\bar\sigma$ play, after the state sequence $\pi$, the corresponding distribution over $E_s$.

    Let us show that the profiles $(\sigma, y)$ in $\Game$ and $(\bar\sigma, y)$ in $\bar\Game$ induce the same distribution over state sequences, up to the first visit to $T$.
    Since the strategy $y$ is memoryless, the mixed action that it plays after a history depends only on the last state of that history, hence only on $\pi$.
    Moreover, the transition function is linear in the mixed action played by Ada, and therefore, conditional on $\pi$, both profiles reach a state $t$ with probability $\delta(s, x_\pi, y(s))(t)$.
    An induction on the length of $\pi$ gives the announced equality of distributions.

    Since the strategy $y$ wins in $\bar\Game$ from every state of $W$, a state of $T$ is reached with probability $1$ under $(\bar\sigma, y)$, and therefore also under $(\sigma, y)$.
\end{proof}

\subsection{Choosing the second mixed actions}

We now choose the second memoryless strategy $p$, whose role will be equivalent to that of the stalling strategy in \Cref{sec:absorbing}.
For a set $X \subseteq S$, we write $\delta(s,x,b)(X) = \sum_{t \in X}\delta(s,x,b)(t)$.
The following lemma gives a technical property of $p$ that we will use.

\begin{lemma} \label{lm:negative_safe_actions}
    There exists a memoryless strategy $p$ with $p(s) \in E_s$ for every $s \notin W$ and $p(s) = p^\circ(s)$ for every $s \in W$, satisfying the following property.
    For every set $R \subseteq S$ with $R \setminus W \neq \emptyset$, there is a nonempty set $C \subseteq R \setminus W$ such that, for every $s \in C$ and $b \in B^s_\crit$,
    \begin{equation} \label{eq:negative_safe_action}
        \delta(s,p^\circ(s),b)(S \setminus R) = 0
        \quad\Longrightarrow\quad
        \delta(s,p(s),b)(C) = 1.
    \end{equation}
\end{lemma}

Thus, whenever $b$ keeps the next state in $R$ against the interior mixed action, it keeps the next state in $C$ against $p(s)$.

\begin{proof}
    We use the following property of finite concurrent reachability games from~\cite[proof of Theorem~2]{CristauDavidHorn2010}, applied to $\bar\Game$.
    There is a mapping $r:S \to \{0,\ldots,|S|\}$, assigning an integer rank to every state, with $r(s)=0$ if and only if $s \in W$.
    For every $s \notin W$, one can choose a vertex $p(s) \in E_s$ such that, for every $b \in B^s_\crit$, at least one of the following holds:
    \begin{enumerate}
        \item every successor $t$ with $\delta(s,p(s),b)(t)>0$ satisfies $r(t)\geq r(s)$;
        \item there exist a vertex $q \in E_s$ and a state $t$ such that $\delta(s,q,b)(t)>0$ and $r(t)>r(s)$.
    \end{enumerate}
    Fix these choices of $p(s)$, and set $p(s)=p^\circ(s)$ on $W$.

    Let $R \subseteq S$ satisfy $R \setminus W \neq \emptyset$, and define:
    $$m = \max_{t \in R} r(t),
    \qquad C = \{t \in R \mid r(t)=m\}.$$
    The set $C$ is nonempty because $R$ is finite.
    Moreover, $m>0$ because $R$ contains a state outside $W$, so $C \subseteq R \setminus W$.

    Let $s \in C$ and $b \in B^s_\crit$, and assume that $\delta(s,p^\circ(s),b)(S \setminus R)=0$.
    By linearity of the transition function, we have:
    $$0 = \sum_{a \in A} p^\circ(s)(a) \cdot \delta(s,a,b)(S \setminus R).$$
    Every term is non-negative, so $\delta(s,a,b)(S \setminus R)=0$ for every $a \in \Supp(p^\circ(s))$.
    By \Cref{lm:negative_critical_relint}, every $q \in P_s$ satisfies $\Supp(q) \subseteq \Supp(p^\circ(s))$.
    Hence, we have $\delta(s,q,b)(S \setminus R)=0$ for every $q \in P_s$: all successors under any vertex of $P_s$, including $p(s)$, belong to $R$.

    Since every state of $R$ has rank at most $m=r(s)$, the second alternative above is impossible.
    The first alternative therefore holds: every successor under $p(s)$ has rank at least $m$.
    Those successors also belong to $R$, so they have rank exactly $m$ and thus belong to $C$.
    This proves $\delta(s,p(s),b)(C)=1$, as required.
\end{proof}

Fix a strategy $p$ as in \Cref{lm:negative_safe_actions}, and let $\sigma$ be the statewise private mixture of $p$ and $p^\circ$.
At every state $s$, both $p(s)$ and $p^\circ(s)$ belong to $P_s$, so every memoryless strategy selected by the initial coins is locally optimal.
To see that the equivalent history-dependent strategy is also locally optimal, let $h$ be a possible history ending at a non-absorbing state $s$, and let $\lambda_h$ be the conditional probability, given $h$, that the coin of $s$ selected $p(s)$.
The mixed action played after $h$ is:
$$\sigma(h) = \lambda_h \cdot p(s) + (1-\lambda_h) \cdot p^\circ(s).$$
It belongs to $P_s$ by convexity.
At histories that cannot occur under the mixture against any strategy of Bertrand, and at absorbing histories, we define $\sigma(h)=p^\circ(\last(h))$, which also belongs to $P_{\last(h)}$.

\begin{lemma} \label{lm:negative_statewise_avoidance}
    From every state outside $W$, the statewise private mixture $\sigma$ satisfies
    $$\prob_{\sigma\tau}(n_T = \infty) > 0$$
    against every critical strategy $\tau$.
\end{lemma}

\begin{proof}
    Fix an initial state outside $W$ and a critical strategy $\tau$.
    Let $R$ consist of $T$ and all states reachable under $(\sigma,\tau)$ before the first visit to $T$.
    Since $\Supp(p(s)) \subseteq \Supp(p^\circ(s))$ at every state, every history compatible with $(\sigma,\tau)$ is also compatible with $(p^\circ,\tau)$.
    Moreover, the initial coin assignment selecting $p^\circ$ everywhere has positive probability.
    Consequently, at every compatible history $\tilde{h}$ avoiding $T$, with $s = \last(\tilde{h})$, every $b \in \Supp(\tau(\tilde{h}))$ satisfies
    $$\delta(s,p^\circ(s),b)(S \setminus R) = 0,$$
    since otherwise a state outside $R$ would be reachable with positive probability under $(\sigma,\tau)$.

    The set $R \setminus W$ contains the initial state, so choose a nonempty $C \subseteq R \setminus W$ as in \Cref{lm:negative_safe_actions}.
    At every compatible history avoiding $T$ and ending at $s \in C$, the mixed action $p(s)$ therefore keeps the next state in $C$ against every action that $\tau$ can play.
    Choose a compatible history $h$ avoiding $T$ that first enters $C$.
    No action has yet been played at a state of $C$, so its coins remain independent and fair conditional on $h$.
    With conditional probability $2^{-|C|}$, all these coins select $p(s)$.
    On that event, the play stays in $C$ forever with probability $1$.
    Since $C \cap T = \emptyset$, we obtain:
    $$\prob_{\sigma\tau}(n_T = \infty)
    \geq \prob_{\sigma\tau}(h) \cdot 2^{-|C|}
    \geq \prob_{\sigma\tau}(h) \cdot 2^{-|S|} > 0.$$
\end{proof}

\subsection{The characterisation}

We can now state our characterisation of games where Ada has a hyperoptimal strategy.

\begin{theorem} \label{thm:negative_statewise_mixture}
    The three following statements are equivalent:
    \begin{enumerate}
        \item\label{it:notinW} we have $s_0 \not\in W$;
        \item\label{it:hyperopt} Ada has a hyperoptimal strategy in $\Game$;
        \item\label{it:pphyperopt} the statewise private mixture of $p$ and $p^\circ$ constructed above is hyperoptimal.
    \end{enumerate}
\end{theorem}

\begin{proof}
    We prove the three implications separately.

    \subp{\ref{it:pphyperopt} implies \ref{it:hyperopt}}
    Trivial.

    \subp{\ref{it:hyperopt} implies \ref{it:notinW}}
    For the sake of contradiction, assume that $s_0 \in W$ and that $\sigma$ is hyperoptimal.
    Let $y$ be the memoryless critical strategy of \Cref{lm:negative_reaching_target}.
    Since $\sigma$ is optimal, \Cref{lm:negative_critical_continuations} gives $\sigma(h) \in P_{\last(h)}$ at every history compatible with $(\sigma,y)$.
    Hence, $T$ is reached almost surely.
    Moreover, criticality gives
    $$\expec_{\sigma y}\bigl(V_{n+1} \mid H_n\bigr) = V_n,$$
    so $\expec_{\sigma y}(V_n) = v_{s_0}$ for every $n$.

    After reaching an absorbing state, the value process is constant and equals the payoff.
    After reaching a state of value $0$, it remains $0$: its conditional expectation is $0$, and all successor values are non-positive.
    In the latter case, only zero-value states can subsequently be visited, so the payoff is also $0$.
    Since $T$ is reached almost surely, we therefore have $V_n \to \mu$ almost surely.
    The variables $V_n$ are uniformly bounded, and dominated convergence gives $\expec_{\sigma y}(\mu) = v_{s_0}$, contradicting hyperoptimality.

    \subp{\ref{it:notinW} implies \ref{it:pphyperopt}}
    Let $\sigma$ be the statewise private mixture of $p$ and $p^\circ$.
    It is locally optimal, hence optimal by \Cref{lm:negative_positive_optimal}.
    Assume towards a contradiction that some strategy $\tau$ satisfies $\expec_{\sigma\tau}(\mu) = v_{s_0}$.
    By \Cref{def:statewise_mixture}, this expectation is the average over the finitely many equally likely initial coin assignments of the expectations of the corresponding memoryless strategies, each of which is at least $v_{s_0}$.
    Thus, all those expectations equal $v_{s_0}$.
    In particular, the component choosing $p^\circ(s)$ at every state has expectation $v_{s_0}$, so $\tau$ is a best response to $p^\circ$.

    By \Cref{lm:negative_relint_best_response}, $\tau$ plays only critical actions along non-absorbing histories compatible with $(p^\circ,\tau)$.
    Since $\Supp(p(s)) \subseteq \Supp(p^\circ(s))$ at every state, every history compatible with $(\sigma,\tau)$ is also compatible with $(p^\circ,\tau)$.
    Complete $\tau$ at incompatible non-absorbing histories by arbitrary critical mixed actions.
    This does not change the probability of any history under $\sigma$, as follows by induction on its length, and makes $\tau$ critical everywhere.
    By \Cref{lm:negative_statewise_avoidance}, we obtain $\prob_{\sigma\tau}(n_T=\infty)>0$.
    Since $s_0 \notin W$ implies $s_0 \notin T$, \Cref{lm:negative_strict_improvement} now gives $\expec_{\sigma\tau}(\mu)>v_{s_0}$, a contradiction.
    Thus, $\sigma$ is hyperoptimal.
\end{proof}

\section{Computational Complexity of the Strict Threshold Problem}\label{sec:complexity}
We now turn the above characterisations into algorithms.
We consider the \emph{strict strategy problem}, for the two classes of games studied above.
The input consists of a finite game $\Game$ and a threshold $\alpha \in \QQ$; all transition probabilities and absorbing rewards are rational numbers, given in binary.
The question is whether
\begin{equation*}
    \exists \sigma\; \forall \tau, \qquad \expec_{\sigma\tau}(\mu) > \alpha.
\end{equation*}
The size of the input includes the binary encodings of all these numbers.
If the initial state is absorbing, the question reduces to comparing its reward with $\alpha$, so we assume below that $s_0$ is non-absorbing.

By the definition of the value, the answer to the strict strategy problem is positive if and only if
\begin{equation} \label{eq:complexity_trichotomy}
    v > \alpha
    \quad\text{or}\quad
    \bigl(v=\alpha \text{ and Ada has a hyperoptimal strategy}\bigr).
\end{equation}
Indeed, when $v>\alpha$, an $\epsilon$-optimal strategy suffices for $0<\epsilon<v-\alpha$; when $v<\alpha$, every strategy has a response with expected payoff below $\alpha$.
The equality case is precisely the definition of hyperoptimality.
In particular, replacing the strict strategy problem with the comparison $v>\alpha$ would omit the equality case.

\subsection{Absorbing games}

\begin{theorem} \label{thm:complexity_absorbing}
    For absorbing games, the strict strategy problem is solvable in polynomial time.
\end{theorem}

\begin{proof}
    We use the notation $\imm$ and $\abs$ from \Cref{sec:absorbing}, and define the threshold constraints
    $$\cons^\alpha_b(x) = \imm(x,b)-\alpha\cdot\abs(x,b),
    \qquad
    P^\alpha = \{x\in\Dist(A)\mid \cons^\alpha_b(x)\geq 0 \text{ for every }b\in B\}.$$
    These constraints have rational coefficients of polynomial encoding length.
    They agree with $\cons_b$ and $P$, respectively, when $\alpha=v$.

    For any strategy profile $(\sigma,\tau)$, let $\zeta$ be its probability of eventual absorption.
    Decomposing the expected payoff according to the round of absorption gives
    \begin{equation} \label{eq:complexity_absorption_identity}
        \expec_{\sigma\tau}(\mu)
        = \alpha\cdot\zeta
        + \sum_{h\in\Hist_{|s_0}} \prob_{\sigma\tau}(h)
        \cdot \cons^\alpha_{\tau(h)}(\sigma(h)).
    \end{equation}
    As usual, the constraint is extended by linearity to a mixed action for Bertrand.
    The series is absolutely convergent: if $M$ bounds the absolute values of the absorbing rewards, then
    $|\cons^\alpha_y(x)|\leq (M+|\alpha|)\cdot\abs(x,y)$, and the corresponding sum of absorption probabilities is $\zeta\leq 1$.

    \subp{Non-negative thresholds}
    Suppose $\alpha\geq 0$.
    We claim that the answer is positive if and only if there exists $x\in\Dist(A)$ such that $\cons^\alpha_b(x)>0$ for every action $b$.
    If such an $x$ exists, then $\abs(x,b)>0$ for every $b$, since zero absorption would also give zero immediate reward and zero constraint.
    The memoryless strategy $x$ therefore ensures absorption almost surely against every strategy $\tau$.
    Writing $\epsilon=\min_b\cons^\alpha_b(x)>0$, the term at the initial history in \eqref{eq:complexity_absorption_identity} is at least $\epsilon$, and all other terms are non-negative.
    Thus, $\expec_{x\tau}(\mu)\geq\alpha+\epsilon$ for every $\tau$.

    Conversely, if no such $x$ exists, the matrix game with entries $\cons^\alpha_b(\d_a)$ has value at most $0$.
    By the minimax theorem, Bertrand has a mixed action $y$ such that $\cons^\alpha_y(x)\leq 0$ for every $x\in\Dist(A)$.
    When Bertrand plays $y$ memorylessly, \eqref{eq:complexity_absorption_identity} gives
    $$\expec_{\sigma y}(\mu)\leq\alpha\cdot\zeta\leq\alpha$$
    for every strategy $\sigma$, proving the claim.
    The condition is decidable by maximising $\epsilon$ subject to $x\in\Dist(A)$ and $\cons^\alpha_b(x)\geq\epsilon$ for every $b$, and testing whether the optimum is positive.

    \subp{Negative thresholds}
    Suppose now $\alpha<0$.
    If $P^\alpha$ is nonempty, define
    $$B^\alpha_\crit = \{b\in B\mid \cons^\alpha_b(x)=0\text{ for every }x\in P^\alpha\}.$$
    We claim that the answer is positive if and only if $P^\alpha$ is nonempty and there exists $p\in P^\alpha$ such that
    \begin{equation} \label{eq:complexity_threshold_stalling}
        \abs(p,b)=0 \qquad\text{for every }b\in B^\alpha_\crit.
    \end{equation}

    First, assume these conditions, and choose $q$ in the relative interior of $P^\alpha$.
    The geometric arguments of \Cref{lm:critical_relint,lm:negative_critical_relint} apply to $P^\alpha$: we have $\Supp(p)\subseteq\Supp(q)$, and $\cons^\alpha_b(q)=0$ exactly for $b\in B^\alpha_\crit$.
    By \eqref{eq:complexity_absorption_identity}, every memoryless strategy in $P^\alpha$ guarantees at least $\alpha$, since $\alpha\cdot\zeta\geq\alpha$.
    Consider the private fair mixture of $p$ and $q$.
    If some response $\tau$ gave expected payoff $\alpha$ against this mixture, it would give expected payoff $\alpha$ against each component.
    Applied to $(q,\tau)$, \eqref{eq:complexity_absorption_identity} would then force $\zeta=1$ and every constraint term at a compatible history to vanish.
    Hence, $\tau$ would play only actions in $B^\alpha_\crit$ along histories compatible with $(q,\tau)$, and therefore also along histories compatible with $(p,\tau)$, by support inclusion.
    By \eqref{eq:complexity_threshold_stalling}, the profile $(p,\tau)$ would never absorb and would have expected payoff $0>\alpha$, a contradiction.
    Thus, the mixture satisfies the strict strategy problem.

    Conversely, assume that the answer is positive.
    If $v=\alpha$, the characterisation for negative-value absorbing games in \Cref{sec:absorbing} gives a stalling strategy, and this is exactly \eqref{eq:complexity_threshold_stalling}, since $P^\alpha=P$ and $B^\alpha_\crit=B_\crit$.
    If $v>\alpha$, the existence of memoryless $\epsilon$-optimal strategies in recursive games~\cite{Everett1957Recursive} gives a memoryless strategy $p$ guaranteeing strictly more than $\alpha$.
    For each action $b$, \Cref{lm:imm_abs} gives either $\abs(p,b)=0$ and $\cons^\alpha_b(p)=0$, or
    $$\frac{\imm(p,b)}{\abs(p,b)}>\alpha,
    \qquad\text{and hence}\qquad \cons^\alpha_b(p)>0.$$
    Therefore, $p\in P^\alpha$, and every $b\in B^\alpha_\crit$ satisfies $\abs(p,b)=0$.
    These two cases exhaust the possibilities by \eqref{eq:complexity_trichotomy}.

    It remains to implement the criterion.
    We first test feasibility of $P^\alpha$ by linear programming.
    If it is nonempty, we identify $B^\alpha_\crit$ by maximising $\cons^\alpha_b(x)$ over $P^\alpha$ for each $b$: the action is critical exactly when this maximum is $0$.
    Finally, we test feasibility of $p\in P^\alpha$ together with the linear equalities \eqref{eq:complexity_threshold_stalling}.
    There are at most $|B|+2$ linear programs, each of polynomial size and with rational coefficients.
    Exact rational linear programming therefore gives the claimed polynomial-time bound.
\end{proof}

\subsection{Negative recursive games}

We write $\exists\RR$ for the class of decision problems reducible in polynomial time to the existential theory of the reals: the satisfiability of existentially quantified Boolean combinations of polynomial equalities and inequalities with integer coefficients.
By Canny's theorem, $\exists\RR\subseteq\PSPACE$~\cite{Canny1988}.

\begin{theorem} \label{thm:complexity_negative}
    For negative recursive games, the strict strategy problem belongs to $\exists\RR$, and hence to $\PSPACE$.
\end{theorem}

We prove this by describing a certificate consisting of two memoryless strategies, a vector of lower bounds on the values, and integer ranks.
This avoids computing the value vector or enumerating the vertices of the polytopes $P_s$.

Let $L=\min(\{0\}\cup\{\mu(s)\mid s\in S_\Abs\})$.
For a vector $u\in[L,0]^S$ satisfying $u_s=\mu(s)$ at absorbing states, define
\begin{align*}
    \cons^u_{s,b}(x)
    &=\sum_{a\in A}x(a)\cdot\left(\sum_{t\in S}\delta(s,a,b)(t)\cdot u_t-u_s\right),\\
    P_s(u)&=\{x\in\Dist(A)\mid\forall b\in B,\ \cons^u_{s,b}(x)\geq 0\}.
\end{align*}
Thus, $\cons^u_{s,b}$ and $P_s(u)$ agree with the constraints and polytopes of \Cref{sec:negative} when $u_s=v_s$ for every $s$.
We also set $T_u=S_\Abs\cup\{s\in S\mid u_s=0\}$, and denote by $n_{T_u}$ its first hitting time.

\begin{lemma} \label{lm:complexity_lower_bounds}
    Let $u$ be as above, and suppose that $\sigma(h)\in P_{\last(h)}(u)$ at every history.
    Then, against every strategy $\tau$, we have $\expec_{\sigma\tau}(\mu)\geq u_{s_0}$.
    If $T_u\neq S$, then, writing $\eta_u=\min_{s\notin T_u}(-u_s)>0$, we have
    \begin{equation} \label{eq:complexity_strict_improvement}
        \expec_{\sigma\tau}(\mu)
        \geq u_{s_0}+\eta_u\cdot\prob_{\sigma\tau}(n_{T_u}=\infty).
    \end{equation}
    Moreover, if $\expec_{\sigma\tau}(\mu)=u_{s_0}$, then $\cons^u_{s,b}(\sigma(h))=0$ for every compatible history $h$ ending at $s$ and every $b\in\Supp(\tau(h))$.
\end{lemma}

\begin{proof}
    The proofs of \Cref{lm:negative_positive_optimal,lm:negative_strict_improvement} use only non-positivity, agreement with the absorbing rewards, and the local constraint inequalities, so they apply to $u$.
    For completeness, let $U_n=u_{S_n}$, and let $\mu_n$ be the absorbing reward if absorption has occurred by round $n$, and $0$ otherwise.
    We have $\mu_n\geq U_n$ and
    $$\expec_{\sigma\tau}(U_{n+1}\mid H_n)
    =U_n+\cons^u_{S_n,\tau(H_n)}(\sigma(H_n))\geq U_n.$$
    Moreover, $\mu_n-U_n\geq\eta_u\cdot\mathbf{1}_{\{n_{T_u}>n\}}$ when $T_u\neq S$.
    Taking expectations and then limits, using bounded convergence for $\mu_n$, proves the first two assertions.

    The expectations $\expec_{\sigma\tau}(U_n)$ are non-decreasing, and their limit is at most $\expec_{\sigma\tau}(\mu)$.
    If this last quantity equals $u_{s_0}$, every increment must be $0$.
    Each increment is an average of non-negative constraint terms.
    Consequently, every such term at a compatible history, with an action in the support of $\tau$, is $0$.
\end{proof}

We then obtain the following characterisation.

\begin{lemma} \label{lm:complexity_certificate}
    The answer to the strict strategy problem is positive if and only if there exist a vector $u$ as above, two memoryless strategies $p,q$, and a mapping $r:S\to\{0,\ldots,|S|\}$ satisfying the following conditions:
    \begin{enumerate}
        \item $p(s),q(s)\in P_s(u)$ for every state $s$;
        \item $\Supp(p(s))\subseteq\Supp(q(s))$ for every state $s$;
        \item $r(s)=0$ for every state $s\in T_u$;
        \item for every non-absorbing state $s$ with $r(s)>0$ and every action $b$ with $\cons^u_{s,b}(q(s))=0$, at least one of the following holds:
        \begin{enumerate}
            \item every $t$ with $\delta(s,p(s),b)(t)>0$ satisfies $r(t)\geq r(s)$;
            \item some $t$ satisfies $\delta(s,q(s),b)(t)>0$ and $r(t)>r(s)$;
        \end{enumerate}
        \item $u_{s_0}\geq\alpha$, and either $u_{s_0}>\alpha$ or $r(s_0)>0$.
    \end{enumerate}
\end{lemma}

\begin{proof}
    By \eqref{eq:complexity_trichotomy}, a positive answer implies $v\geq\alpha$.

    \subp{From a strategy to a certificate}
    Suppose that the answer is positive, and set $u_s=v_s$ for every state $s$.
    If $v>\alpha$, choose a locally optimal memoryless strategy $p$, set $q=p$, and give every state rank $0$.
    Such a strategy exists by \Cref{lm:negative_positive_optimal}, and all five conditions hold.

    If $v=\alpha$, Ada has a hyperoptimal strategy.
    By \Cref{thm:negative_statewise_mixture}, the initial state lies outside $W$.
    Choose $q=p^\circ$, and take $p$ and $r$ from the proof of \Cref{lm:negative_safe_actions}.
    Conditions 1--3 hold by construction and by \Cref{lm:negative_critical_relint}, since $T_u=T\subseteq W$.
    At a state of positive rank, the actions $b$ with $\cons^u_{s,b}(q(s))=0$ are exactly the critical actions.
    The first alternative in that proof is condition 4(a).
    For its second alternative, every vertex of $P_s$ has support contained in $\Supp(q(s))$, so a successor of higher rank reachable under a vertex is also reachable under $q(s)$.
    This gives condition 4(b).
    Finally, $r(s_0)>0$, so condition 5 holds.

    \subp{From a certificate to a strategy}
    Let $u,p,q,r$ satisfy the five conditions, and let $\sigma$ be the statewise private mixture of $p$ and $q$.
    Every memoryless component of $\sigma$ satisfies the local inequalities for $u$.
    The equivalent history-dependent strategy does as well, since its mixed action at $s$ is a convex combination of $p(s)$ and $q(s)$.
    By \Cref{lm:complexity_lower_bounds}, all these strategies guarantee at least $u_{s_0}$.
    Thus, if $u_{s_0}>\alpha$, the conclusion follows.

    Suppose now $u_{s_0}=\alpha$, so $r(s_0)>0$.
    Assume for a contradiction that some strategy $\tau$ satisfies $\expec_{\sigma\tau}(\mu)=\alpha$.
    Since every component guarantees at least $\alpha$, and the component choosing $q$ at every state has positive probability, we also have $\expec_{q\tau}(\mu)=\alpha$.
    By \Cref{lm:complexity_lower_bounds}, every action $b$ played by $\tau$ at a history compatible with $(q,\tau)$ satisfies $\cons^u_{s,b}(q(s))=0$, where $s$ is the last state.
    By condition 2, every history compatible with $(\sigma,\tau)$ is also compatible with $(q,\tau)$, so this conclusion holds along the mixture as well.

    Let $R$ consist of $T_u$ and all states reachable under $(\sigma,\tau)$ before the first visit to $T_u$.
    Define
    $$m=\max_{t\in R}r(t),\qquad C=\{t\in R\mid r(t)=m\}.$$
    Since $s_0\in R$ and $r(s_0)>0$, the set $C$ is nonempty and disjoint from $T_u$.
    At every compatible history avoiding $T_u$ and ending at $s$, every action $b$ in the support of $\tau$ satisfies
    $$\delta(s,q(s),b)(S\setminus R)=0.$$
    Indeed, the all-$q$ component has positive probability and is compatible with that history, so a successor outside $R$ would contradict the definition of $R$.

    Suppose in addition that $s\in C$.
    Condition 4(b) is impossible, since all successors under $q(s)$ belong to $R$ and have rank at most $m=r(s)$.
    Hence, condition 4(a) holds.
    By condition 2, every successor under $p(s)$ also belongs to $R$; its rank is therefore exactly $m$, so it belongs to $C$.
    Thus, playing $p(s)$ keeps the next state in $C$ at every such history.

    Choose a compatible history $h$ avoiding $T_u$ that first enters $C$.
    No action has yet been played in $C$, so, conditional on $h$, its private coins are still independent and fair.
    With probability $2^{-|C|}$, they all choose $p(s)$; on that event, the play remains in $C$ forever with probability $1$.
    Consequently,
    $$\prob_{\sigma\tau}(n_{T_u}=\infty)
    \geq\prob_{\sigma\tau}(h)\cdot 2^{-|C|}>0.$$
    By \eqref{eq:complexity_strict_improvement}, this implies $\expec_{\sigma\tau}(\mu)>u_{s_0}=\alpha$, a contradiction.
    Hence, $\sigma$ witnesses a positive instance of the strict strategy problem.
\end{proof}

We can then prove our complexity result.

\begin{proof}[Proof of \Cref{thm:complexity_negative}]
    We encode the certificate of \Cref{lm:complexity_certificate} by an existential formula over the reals.
    Its variables are $u_s$, the probabilities $p(s)(a)$ and $q(s)(a)$, and one variable $r(s)$ for each state.
    Membership in $\Dist(A)$ is expressed by non-negativity and the equation that the probabilities sum to $1$.
    The constraints $\cons^u_{s,b}(p(s))\geq 0$ and $\cons^u_{s,b}(q(s))\geq 0$ have degree at most $2$.
    Support inclusion is expressed by
    $$p(s)(a)=0\quad\lor\quad q(s)(a)>0,$$
    and the rank domain by $\bigvee_{j=0}^{|S|}r(s)=j$.
    Condition 3 uses the implications $u_s=0\Rightarrow r(s)=0$, together with $r(s)=0$ at absorbing states.

    For condition 4, all quantifiers range over explicitly given finite sets and are expanded into conjunctions or disjunctions.
    Its two alternatives are expressed, respectively, by
    \begin{align*}
        &\bigwedge_{t\in S}\bigl(\delta(s,p(s),b)(t)=0\ \lor\ r(t)\geq r(s)\bigr),\\
        &\bigvee_{t\in S}\bigl(\delta(s,q(s),b)(t)>0\ \land\ r(t)>r(s)\bigr).
    \end{align*}
    The antecedent and condition 5 are also Boolean combinations of polynomial inequalities.
    Thus, the entire formula has polynomial size, polynomially many variables, and degree at most $2$.
    All coefficients are rational and have polynomial encoding length; multiplying each atomic polynomial by a positive common denominator gives integer coefficients with polynomial encoding length.
    The formula is satisfiable exactly when the answer is positive, by \Cref{lm:complexity_certificate}.
    This proves membership in $\exists\RR$, and the $\PSPACE$ bound follows from~\cite{Canny1988}.
\end{proof}

\subsection{A lower bound for negative recursive games}

The upper bound above does not give a polynomial-time algorithm for negative recursive games.
The following reduction shows that even a restricted subclass captures the square-root-sum problem: given positive integers $a_1,\ldots,a_n,k$ in binary, decide whether $\sum_{i=1}^n\sqrt{a_i}>k$.
The strict and non-strict versions are polynomial-time equivalent: equality with the integer $k$ is possible only when all the $a_i$ are perfect squares, and this case can be decided in polynomial time.

\begin{theorem} \label{thm:complexity_hardness}
    The strict strategy problem for negative recursive games is square-root-sum-hard under polynomial-time many-one reductions.
    This holds already for safety games in which an absorbing state is reached almost surely under every strategy profile.
\end{theorem}

\begin{proof}
    We first remove perfect-square summands and subtract their square roots from $k$.
    Instances with no remaining summands, or with a non-positive residual threshold, are decided directly and mapped to fixed positive or negative instances.
    For the remaining instances, we use the construction of Etessami and Yannakakis~\cite[proof of Theorem~5.1]{EtessamiYannakakis2008}.
    Given a square-root-sum instance, it produces in polynomial time a finite concurrent reachability game with absorbing states $t,z$ whose value for reaching $t$ is
    $$V=D+E\cdot\sum_{i=1}^n\sqrt{a_i},$$
    where $D,E$ are polynomial-time computable rationals and $E>0$.
    Set $\beta=D+E k$.

    We also use the fact that this construction absorbs almost surely under every strategy profile.
    Indeed, in each of its gadgets, a visit to the concurrent state is followed by absorption within two transitions with probability at least $1-c_2$, where $0\leq c_2<1/4$ is the gadget's return probability.
    The only state preceding the gadgets makes a single probabilistic choice among them.
    Thus, the probability of avoiding both absorbing states tends to $0$ under every profile.

    On the same arena, give $t$ reward $0$ and $z$ reward $-1$.
    This is a safety game for Ada, with $z$ as the unsafe state.
    Since absorption is almost sure under every profile, we have
    $$\expec_{\sigma\tau}(\mu)
    =-\prob_{\sigma\tau}(\text{reach }z)
    =\prob_{\sigma\tau}(\text{reach }t)-1.$$
    The resulting negative recursive game therefore has value $v=V-1$.

    Bertrand has an optimal memoryless strategy in this game.
    To see this, choose at each state an optimal mixed action for Bertrand in the matrix game with successor values, as in \Cref{lm:negative_positive_nonempty}.
    Against this strategy, $(v_{S_n})_{n\geq 0}$ is a bounded supermartingale under every strategy of Ada.
    Since absorption is almost sure, $v_{S_n}$ converges almost surely to the payoff; bounded convergence gives expected payoff at most $v$.
    In particular, Ada has no hyperoptimal strategy.

    Set $\alpha=\beta-1$.
    By \eqref{eq:complexity_trichotomy}, the resulting instance has a positive answer exactly when
    $$v>\alpha
    \quad\Longleftrightarrow\quad V>\beta
    \quad\Longleftrightarrow\quad\sum_{i=1}^n\sqrt{a_i}>k.$$
    The arena and all rational numbers have polynomial encoding length, completing the reduction.
\end{proof}


\bibliography{bibli}

\end{document}